\documentclass[aps,prb,twocolumn,amsmath,amssymb,superscriptaddress,scrartcl,eqsecnum,longbibliography]{revtex4-1}
\usepackage[T1]{fontenc}
\usepackage[latin9]{inputenc}
\usepackage{color}
\usepackage{verbatim}
\usepackage{amsmath}
\usepackage{amssymb}
\usepackage{graphicx}

\usepackage{tikz}
\usepackage{diagbox}
\usepackage[normalem]{ulem}

\makeatletter
\PassOptionsToPackage{caption=false}{subfig} 
\usepackage{hyperref}
\hypersetup{
breaklinks=true,
colorlinks=true,
citecolor=blue,
linkcolor=black,
filecolor=black,
urlcolor=black
}
\IfFileExists{lmodern.sty}{\usepackage{lmodern}}{}

\renewcommand{\bar}{\overline}
\renewcommand{\tilde}{\widetilde}

\newcommand{\SU}{\operatorname{SU}}

\newcommand{\SL}{\operatorname{SL}}

\newcommand{\calC}{\mathcal{C}}

\makeatletter

\newcommand*{\wideboxed}[1]{\setlength{\fboxsep}{1ex}%
  \fbox{\m@th$\displaystyle#1$}}
\makeatother

\def\be{\begin{equation}}
\def\ee{\end{equation}}

\def\b{\boldsymbol}

\makeatother

\begin{document}

\title{Classification of $\SL_2$ 
deformed Floquet Conformal Field Theories}

\author{Bo Han}

\affiliation{Theory of Condensed Matter Group, Cavendish Laboratory, University of Cambridge  \\
J.~J.~Thomson Avenue, Cambridge CB3 0HE, United Kingdom}

\author{Xueda Wen}

\affiliation{Department of Physics, Massachusetts Institute of Technology, Cambridge, MA 02139, USA}

\begin{abstract}
Classification of the non-equilibrium quantum many-body dynamics is a challenging problem in condensed matter
physics and statistical mechanics.
In this work, we study the basic question that whether a (1+1) dimensional
conformal field theory (CFT) is stable or not under
a periodic driving with $N$ non-commuting Hamiltonians.
Previous works showed that a Floquet (or periodically driven)  CFT
driven by certain $\SL_2$ deformed Hamiltonians exhibit both non-heating
(stable) and heating (unstable) phases. 
In this work, we show that the phase diagram depends on the types of driving Hamiltonians.
In general, the heating phase is generic, but the non-heating 
phase may be absent in the phase diagram.
For the existence of the non-heating phases, we give sufficient and necessary conditions
for $N=2$, and sufficient conditions for $N>2$.
These conditions are composed of $N$ layers of data, with each layer 
determined by the types of driving Hamiltonians.
Our results also apply to the single quantum quench problem with $N=1$.

\end{abstract}
\maketitle

%\tableofcontents

%\newpage

\section{Introduction}

Non-equilibrium many-body dynamics have received extensive attention recently because they show exotic properties that are missing in static systems and also they can be realized in experiments 
such as optical lattice and cold atomic systems. For example, 
a periodic drive creates novel systems that may not have an equilibrium analog, such as 
Floquet topological 
phases\cite{Jiang:2011xv,demler2010prb,rudner2013anomalous,else2016prb,ashvin2016prx,roy2016prb,Po:2016qlt,roy2017prb,roy2017prl,po2017radical,yao2017floquetspt,po2017timeglide,ashvin2019prb,
glorioso2019effective} 
and time crystals\cite{khemani2016phase, else2016timecrystal, sondhi2016phaseI, sonhdi2016phaseII, Else:2017ghz, normal2017timecrystal, lukin2017exp, zhang2017observation, normal2018classical}. 
Moreover, a periodic drive is also one of the basic protocols to study non-equilibrium phenomena, such as localization-thermalization transitions, prethermalization,
dynamical Casimir effect, etc. 
\cite{rigol2014prx, abanin2014mbl,abanin2014theory, abanin2015prl, abanin2015rigorous, abanin2015effectiveh,
Law1994,Dodonov1996,martin2019floquet}

Although there are rich properties and applications in 
the time-dependent driving physics in quantum many-body systems, 
exactly solvable setups are very rare. In general, we need to resort to numerical methods 
that are limited to a small-system size. 
In this work, we are interested in a quantum $(1+1)$ dimensional conformal field theory (CFT), which may be
viewed as the low energy effective field theory of a many-body system at the critical point. 
For $(1+1)$D CFTs,
the property of conformal invariance can be exploited to  constrain the operator content
of the critical theory,\cite{belavin1984infinite,Shenker1984,francesco2012conformal}
which makes it tractable for the study of non-equilibrium dynamics, such as the quantum quench problems.
\cite{Calabrese_2005,CC2006}
For a time-dependent driven CFT, however, relatively little is known on the analytical
properties of the non-equilibrium dynamcis.

Recent study along this direction was initialized in Ref.~\onlinecite{Wen:2018vux,wen2018floquet}, where two non-commuting Hamiltonians are used to drive the CFT periodically in time.
One of the driving Hamiltonians is chosen as the uniform one, and the other is chosen by
deforming the uniform one with a sine-square deformation (SSD),\cite{Nishino2011prb,2011freefermionssd,katsura2012sine,
ishibashi2015infinite,ishibashi2016dipolar,
Okunishi:2016zat,Wen:2016inm,Tamura:2017vbx,Tada:2017wul,Wen:2018vux,caputa2020geometry,liu2020analysis,tada2019time,Ryu2019SSD} 
which we will introduce in detail shortly.
In this periodically driven CFT (or Floquet CFT),
it is found there are both heating and non-heating phases, separated by a 
critical (phase transition) line.
One of the `order parameters'  characterizing the phase diagram 
is the time evolution of entanglement entropy. It was found that the entanglement entropy
 grows linearly in time in the heating phase, oscillates in the non-heating phase, 
 and grows logarithmically in time at the phase transition.\cite{wen2018floquet}

 Later in Ref.~\onlinecite{fan2019emergent}, further interesting features were found in the same
 setup. For example, the total energy grows exponentially in time in the heating phase,
 oscillates in the non-heating phase, and grows polynomially in time at the phase transition.
In particular, in the heating phase, 
 it was found there are interesting emergent spatial structures during the driving:
 The (chiral and anti-chiral) energy-momentum densities form an array of `peaks' in the real space (See also Ref.~\onlinecite{Zurich2019} for the study of the energy density distribution).  
 In Ref.~\onlinecite{fan2019emergent}, it was also found that 
 the entanglement pattern in a Floquet CFT is closely related to the energy-momentum
 density distributions. The main contribution of quantum entanglement in the Floquet CFT
 comes from those between nearby energy-momentum density peaks of the same chirality.

Most recently, the types of driving sequences are generalized from periodic to 
quasi-periodic~\cite{QuasiPeriodic,lapierre2020fine} and random ones.~\cite{RandomCFT}
On the one hand, it was found that the heating phase is generic
in all these three types of driving sequences. In particular, 
the features  as found in the periodic driving in 
Ref.~\onlinecite{fan2019emergent} turn out to be also generic in the other two types of drivings.
On the other hand, there are some new features in the phase diagrams of 
the quasi-periodic and random drivings.
For example, in a quasi-periodically driven CFT with Fibonacci sequence, 
the non-heating phases form a Cantor set of measure zero, and the heating rates in the
heating phases exhibit self-similarity structures.
In the random driving, the driven CFT is generally in the heating phase,
but with some isolated exceptional points (See Ref.~\onlinecite{RandomCFT} for more details).
The mechanism of the heating phase 
in a randomly driven CFT is analogous to the Anderson localization 
in (1+1)d disordered system.~\cite{RandomCFT}
In short, the phase diagrams of time-dependent driven CFTs 
depend on the types of driving sequences.

In this work, by fixing the driving sequence to be a periodic one, 
we are interested in how the types of driving Hamiltonians affect the 
phase diagrams of a Floquet CFT. Let us specify the meaning of  
``Hamiltonian types'' first.
Considering a CFT defined on a circle of length $L$, 
the driving Hamiltonians we consider are of the following form:
\be\label{Hcft}
H_{\text{CFT}}=H_{\text{chiral}}+H_{\text{anti-chiral}},
\ee
i.e., one can decompose the total Hamiltonian as the sum of chiral and anti-chiral parts.
In terms of the energy-momentum tensor $T(x)$, we have
\be
\label{Hcfti}
H_{\text{chiral}}=\frac{1}{2\pi}\int_{0}^{L} f(x)\, T (x) dx,
\ee
where $f(x)$ is an envelope function which deforms the chiral energy-momentum density, 
and it is similar for the anti-chiral part, where the anti-chiral energy-momentum tensor is denoted as $\bar{T}(x)$. 
Here $T(x)$ and $\bar{T}(x)$ are related to the energy density $T_{00}$
and momentum density $T_{01}(x)$ as $T_{00}=\frac{1}{2\pi}(T+\overline{T})$ and
$T_{01}=\frac{1}{2\pi}(T-\overline{T})$.
For different driving Hamiltonians, we can choose
different envelope functions $f(x)$.
In this work, we are interested in the deformation of a single wavelength, with
\be
\label{fx_SL2}
f(x)=
\sigma^0+\sigma^+\cos\frac{2\pi q x}{L}+\sigma^-\sin\frac{2\pi q x}{L},\quad q\in \mathbb Z,
\ee
where $\sigma^0$ and $\sigma^{\pm}$ are real numbers which characterize the deformation.
With this deformation, the Hamiltonian can be written as
\be\label{Hdeform_SL2}
H_{\text{chiral}}=\frac{2\pi}{L}\left(\sigma^0 L_0+\sigma^+ L_{q,+}
+\sigma^- L_{q,-}
\right)-\frac{\pi c}{12 L}, 
\ee
where we have defined
$L_{q,+}:=\frac{1}{2}(L_q+L_{-q})$ and $L_{q,-}:=\frac{1}{2i}(L_q-L_{-q})$,
with $L_{n}:=\frac{c}{24}\delta_{n,0}+\frac{L}{2\pi}\int_0^L \frac{dx}{2\pi}\, e^{i\frac{2\pi n}{L}x} \, T(x)
$ being the generators of Virasoro algebra,~\cite{francesco2012conformal}
and $c$ is the central charge.
One can find that $H_{\text{chiral}}$ in \eqref{Hdeform_SL2}
 is composed of three generators that generate the $\SL^{(q)}(2,\mathbb R)$ group, which is isomorphic to the $q$-fold 
cover of $\SL(2,\mathbb R)$\cite{witten1988}.
For this reason, we call the periodically driven CFT with the driving Hamiltonians in 
\eqref{Hcft}-\eqref{fx_SL2} as the $\SL_2$ deformed Floquet CFT.
In general, the Hamiltonians in \eqref{Hdeform_SL2} can be classified into three types
based on the quadratic Casimir:~\cite{ishibashi2015infinite,ishibashi2016dipolar,tada2019time}
\begin{align}\label{Eq:Casimir}
c^{(2)}&:= -(\sigma^0)^2 + (\sigma^+)^2 + (\sigma^-)^2.
\end{align}
Different types of Hamiltonians can be classified as follows:
\begin{equation}
\label{3Types}
\small
\def\arraystretch{1.5}
\begin{tabular}{| l|l|l|l |}
\hline
    Quadratic Casimir  & $c^{(2)}<0$  & $c^{(2)}=0$ & $c^{(2)}>0$\\
    \hline
    Hamiltonian Type & Elliptic    & Parabolic & Hyperbolic \\
    \hline
\end{tabular}
\end{equation}

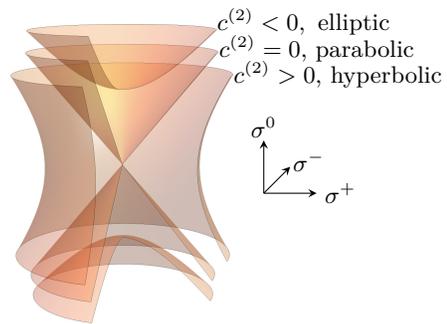
\begin{figure}[t]
\centering
\begin{tikzpicture}
\node[inner sep=0pt] (russell) at (0pt,0pt)
    {\includegraphics[width=.2\textwidth]{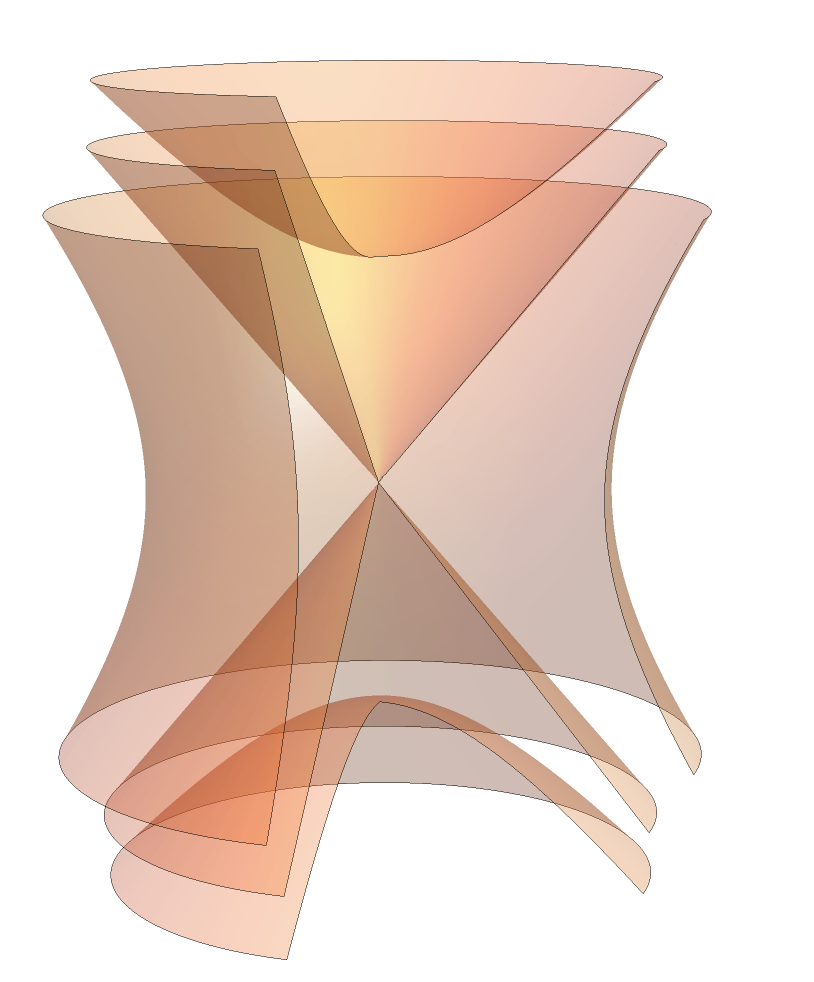}};
    \node at (65pt, 55pt){$c^{(2)}<0,\ \text{elliptic}$};
        \node at (70pt, 45pt){$c^{(2)}=0,\,\text{parabolic}$};
          \node at (78pt, 35pt){$c^{(2)}>0,\,\text{hyperbolic}$};
          
          \draw [>=stealth,->] (50pt, -10pt)--(70pt,-10pt);
          \draw [>=stealth,->] (50pt, -10pt)--(50pt,10pt);
           \draw [>=stealth,->] (50pt, -10pt)--(60pt,0pt);
                     \node at (79pt, -10pt){$\sigma^+$};
                     \node at (50pt, 15pt){$\sigma^0$};
                     \node at (67pt, 2pt){$\sigma^-$};
\end{tikzpicture}
\caption{Different types of manifolds determined by Eq.\eqref{Eq:Casimir} with different quadratic Casimir $c^{(2)}$. Each single point on the manifold specifies a deformed Hamiltonian through
\eqref{Hcfti} and \eqref{fx_SL2}.
Any point on the manifold is $\SL(2,\mathbb R)$ equivalent to arbitrary points on the \textit{same} manifold.
} 
\label{Casimir}
\end{figure}

The reason we use these terminologies will be clear when we discuss operator evolutions in Sec.~\ref{Sec:FloquetCFT}.
Therein, we will see that the operator evolutions governed by different Hamiltonian 
types correspond to the M\"obius transformations of hyperbolic/parabolic/elliptic types.
An intuitive picture to understand the difference of these three Hamiltonian types
 is to consider the manifold determined by \eqref{Eq:Casimir}
with different signs of $c^{(2)}$, as shown in Fig.~\ref{Casimir}.
Recently, the properties of energy spectrum for different types of Hamtiltonians
have been studied in literature.~\cite{Okunishi:2016zat,Wen:2016inm,ishibashi2015infinite,ishibashi2016dipolar,tada2019time}
As a remark, the specific choice of $q=1$, $\sigma^0=-\sigma^+=1/2$ and
$\sigma^-=0$ in \eqref{fx_SL2} corresponds to the CFT with SSD, as used in the simplest setup
of a Floquet CFT in Ref.~\onlinecite{wen2018floquet}.

Our motivation in this work is to classify the non-equilibrium dynamics 
in a Floquet CFT with all possible choices of Hamiltonian types in \eqref{3Types}.
More specifically, during the periodic driving, we choose $N$ ($N\ge 2$) 
non-commuting Hamiltonians, with each Hamiltonian specified
by a certain $(\sigma^0,\, \sigma^+,\,\sigma^-)$ in \eqref{Hdeform_SL2}.
What we mean by `classification' is to determine whether
 there are heating phases/non-heating phases/phase transitions
in the phase diagram.
The main results we found can be summarized as follows.
For arbitrary choices of Hamiltonian types, there are always heating phases in 
the Floquet CFT, i.e., the heating phase is generic.
The non-heating phases, however, do not always exist in the phase diagram.
When there are $N=2$ driving Hamiltonians, we give the sufficient and necessary conditions for the
existence of non-heating phases in the phase diagram, as shown in 
Table.~\ref{classify}.
For $N>2$, we give sufficient conditions for the existence of non-heating phases, as
presented in Sec.~\ref{Sec:General_N}. 
These conditions can be summarized as follows:
There are in total $N$ layers of data.
At each layer $n$, we pick arbitrary $n$ driving Hamiltonians by keeping
the time order of driving. If there is at least one elliptic Hamiltonian in $\{H_{i_1}, \,H_{i_2},\cdots, H_{i_n}\}$, then there must exist non-heating phases.
If all the Hamiltonians in $\{H_{i_1}, \,H_{i_2},\cdots, H_{i_n}\}$ are non-elliptic (either 
parabolic or hyperbolic),
then the following condition ensures the existence of non-heating phases:
\be
\label{eta_N_intro}
\wideboxed{
\exists \ \eta_n<0,\quad n=1, 2,\cdots, N
}
\ee
where $\eta_n$ is the indicator as defined in \eqref{eta_N}.
It is noted that  $\eta_n$ is only determined by the vectors 
$(\sigma^0_i,\,\sigma^+_i,\,\sigma^-_i)$ that characterize
these driving Hamiltonians.
One can find there are in total $2^N-1$ conditions for $N$ driving Hamiltonians.
If at least one of these conditions is satisfied, then there must exist non-heating
phases in the phase diagram.
We suspect that these conditions are also necessary, 
i.e., if there exists a non-heating phase in the phase diagram, then at least one
of the $2^N-1$ conditions mentioned above should be satisfied, 
although we have not yet found a mathematical proof when $N>2$.

\bigskip

The structure of this paper is organized as follows.
In Sec.~\ref{Sec:FloquetCFT}, 
we introduce the setup for a general Floquet CFT with $\SL_2$ deformations.
Then we study the operator
evolution corresponding to different Hamiltonian types, based on which the entanglement/energy-momentum evolution can be obtained.
In Sec.~\ref{Sec:ClassifyFloquetCFT}, we study how the phase diagrams depend on the
types of $N$ driving Hamiltonians in a Floquet CFT. We give sufficient conditions 
for the existence of non-heating phases for arbitrary $N$, 
and illustrate these conditions with examples $N=1$, $2$, and $3$.
We give some discussions and conclude in Sec.~\ref{Sec:Conclusion}.
There are also several appendices focusing on the detailed features of the phase
diagrams with different types of driving Hamiltonians.

\section{$\SL_2$  deformed Floquet CFT}
\label{Sec:FloquetCFT}

\subsection{Setup}
\label{Sec:Setup}

The setup we consider is based on  a (1+1) dimensional CFT  on a circle of length $L$ 
with periodic boundary conditions. 
The driving Hamiltonians we choose are of the form in Eqs. \eqref{Hcft}-\eqref{fx_SL2}.
Then we study a time-dependent driving with a discrete and periodic sequence:
\be\label{DiscreteDriving}
|\Psi_{n}\rangle= \big(U_N\cdots U_2\cdot U_1\big)^m |\Psi_0\rangle,\quad \text{with}\quad U_j=e^{-iH_j T_j},
\ee
where $n=Nm$, and $T_j$ is the time duration of driving with Hamiltonian $H_j$. 
Within each period, there are in general $N$ (possibly) different driving Hamiltonians.
Here the initial state $|\Psi_0\rangle$ may be taken as the ground state 
of a uniform CFT Hamiltonian 
\be
\label{H0}
H_0=\frac{1}{2\pi}\int_{0}^{L} T (x) dx+\text{anti-chiral part}.
\ee
It is noted that the initial state can also be chosen as an excited state or even
a thermal ensemble at finite temperature.~\cite{Thermal}
Each driving Hamiltonian $H_i$ in \eqref{DiscreteDriving} 
can be chosen with different envelope functions $f_i(x)$ in \eqref{fx_SL2}, 
which are characterized by a triple $(\sigma^0_i, \sigma^+_i,\sigma^-_i)$.
Our goal is to classify the non-equilibrium dynamics and phase diagrams with respect to the types 
of $H_i$ ($i=1,\cdots, N$) in the $N$ dimensional 
parameter space spanned by 
\be
\small
\left\{\left(\frac{T_1}{l}, \cdots, \frac{T_N}{l}\right)\Big| 0< \frac{T_i}{l}<\infty, i=1, \cdots, N \right\},
\ee
where $l:=L/q$ is the wavelength of deformation in $f(x)$ [see Eq.\eqref{fx_SL2}].

As a remark, in the above discussions, we only specify the 
deformation of the chiral part of the Hamiltonian, which are characterized 
by the vector $(\sigma^0_i, \sigma^+_i,\sigma^-_i)$. 
The deformation of the anti-chiral parts are characterized by another 
independent vector $(\sigma^0_i\,', \sigma^+_i\,',\sigma^-_i\,')$,
because we choose periodic boundary conditions and 
the chiral and anti-chiral modes are decoupled from each other.
In the following study, without loss of generality, we will focus 
on the deformation of the chiral parts. The analysis of the anti-chiral 
parts can be performed in the same way.

\subsection{Hamiltonian types and operator evolution}
\label{Sec:HamiltonianType}

To understand the non-equilibrium dynamics 
under the driving in Eq.~\eqref{DiscreteDriving}, we first study the operator 
evolution, based on which the time evolution of correlation functions
such as the entanglement entropy and the energy-momentum density can be obtained
\cite{Wen:2018vux,wen2018floquet,fan2019emergent,Zurich2019}.

\begin{figure}[t]
\centering
\begin{tikzpicture}
\draw  [xshift=-40pt][thick](20pt,-40pt)--(20pt,40pt);
\draw  [xshift=-40pt][thick](70pt,-40pt)--(70pt,40pt);

\draw [xshift=-40pt][>=stealth,->] (35pt, 5pt)--(35pt,20pt);
\draw [xshift=-40pt][>=stealth,->] (35pt, 5pt)--(50pt,5pt);
%\draw [ >=stealth,<-] (10pt, -20pt)--(10pt,0pt); \draw [>=stealth,->](10pt,0pt)--(10pt,20pt);

\draw [xshift=-40pt](55pt, 30pt)--(55pt,40pt);
\draw [xshift=-40pt](55pt, 30pt)--(65pt,30pt);

\node at (55-40pt,5pt){$x$};
\node at (40-40pt,20pt){$\tau$};
\node at (60-40pt,35pt){$w$};

\draw [dashed](-20pt, -9.8pt)--(30pt,-9.8pt);
\node at (60-40pt,-10pt){$\bullet$};
\node at (50-40pt,-19pt){$\mathcal{O}(x,\tau)$};

\node at (20-40pt,-45pt){$x=0$};
\node at (70-40pt,-45pt){$x=L$};

\end{tikzpicture}
\caption{
Path integral representation of the correlation function $\langle G|\mathcal{O}(x,\tau)|G\rangle$
in a CFT with periodical boundary conditions.
Here  $x=0$ and $x=L$ are identified.}
\label{ConformalMapQ}
\end{figure}
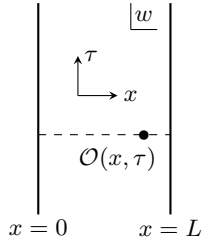

Let us start from driving the CFT with a single Hamiltonian $H$ 
with a time duration $t$.
We consider the method as used in Ref.~\onlinecite{wen2018floquet}, which we briefly
sketch as follows. In the Euclidean spacetime,
the correlation function $\langle \Psi(t)|\mathcal{O}(x)|\Psi(t)\rangle=\langle G|e^{H\tau}\mathcal{O}(x)
e^{-H\tau}|G\rangle=\langle G|\mathcal{O}(x,\tau)|G\rangle$, where $\tau=it$,
 can be considered as the path integral on a $w$-cylinder with the operator $\mathcal{O}$ inserted
 at $(x,\tau)$, as depicted in Fig.~\ref{ConformalMapQ}.
This cylinder can be mapped to a $q$-sheet Riemann surface with a conformal map $z=e^{i\frac{2\pi q w}{L}}$.
Then the Hamiltonian in Eqs.\eqref{Hcft}-\eqref{fx_SL2}
can be written as $H=H^{(z)}+\bar{H}^{(\bar{z})}$, where
\be
\begin{split}
H^{(z)}
=\frac{2\pi}{l}
\oint
&\frac{dz}{2\pi i}
\Big[
\sigma^0 z+\frac{1}{2}(\sigma^+-i\sigma^-)z^2\\
&+\frac{1}{2}(\sigma^++i\sigma^-)
\Big]T(z)
-\frac{\sigma^0 \pi c}{12 l}.
\end{split}
\ee
One can further perform a M\"obius transformation
$z=(\mathfrak{a}\tilde{z}+\mathfrak{b})/(\mathfrak{c}\tilde{z}+\mathfrak{d})$, where $\mathfrak{a}=-\frac{i}{2}$,
$\mathfrak{b}=\frac{\sigma^++i\sigma^-}{\sqrt{c^{(2)}}}$, $\mathfrak{c}=
\frac{-\sqrt{c^{(2)}}+i\sigma^0}{2(\sigma^++i\sigma^-)}$, and 
$\mathfrak{d}=-\frac{\sigma^0+i\sqrt{c^{(2)}}}{\sqrt{c^{(2)}}}$.
Then the Hamiltonian defined on the $q$-sheet $\tilde{z}$ Riemann surface
is of the simple form:
\be
\label{HtildeZ}
H^{( \tilde{z} )}=-\frac{2\pi i \sqrt{c^{(2)}}}{l}
\oint\frac{d\tilde{z}}{2\pi i} \,\tilde{z}\,T(\tilde{z})
-\frac{\sigma^0 \pi c}{12 l}.
\ee

Several remarks here. First, from \eqref{HtildeZ}, one can already
see the difference for positive and negative $c^{(2)}$.
The choice of branch cut of $\sqrt{\cdots}$ will not affect the
results of operator evolution as presented later in
Eqs.\eqref{EllipticMobius}, \eqref{ParabolicMobius}, and \eqref{HyperbolicMobius}.
Second, for $c^{(2)}=0$, the expression in \eqref{HtildeZ} is not well defined.
To obtain the operator evolution, one can do the calculation by keeping nonzero $c^{(2)}$ and 
 take the limit $c^{(2)}\to 0$ for either $c^{(2)}>0$ or $c^{(2)}<0$ in the 
 last step.\cite{Okunishi:2016zat,Wen:2016inm,wen2018floquet,Han2020appearsoon}

On the $q$-sheet $\tilde{z}$ Riemann surface, the operator
evolution becomes a dilatation:
$e^{H^{(\tilde{z})}\tau}\mathcal{O}(\tilde{z},\bar{\tilde{z}})
e^{-H^{(\tilde{z})}\tau}=\lambda^h \mathcal{O}
(\lambda \tilde{z}, \bar{\tilde{z}})$, where
$\lambda=e^{-\frac{2\pi i \sqrt{c^{(2)}}}{l}\tau}$, and $h$ is the conformal
dimension of the operator $\mathcal{O}$.
Then by mapping back to the $z$-surface, one can find 
the operator evolves as 
\be\label{OPevolutionAppendix}
e^{H^{(z)}\tau}\mathcal{O}(z,\bar{z})
e^{-H^{(z)}\tau}=
\left(\frac{\partial z'}{\partial z}\right)^h
\mathcal{O}(z',\bar{z}'). 
\ee
By doing an analytical continuation $\tau=it$, where $t$ the time duration of
driving, one has
\be\label{Zmob}
z'=\frac{\alpha z+\beta}{\beta^* z+\alpha^*}
=: M \cdot z, \quad M= \begin{pmatrix}
\alpha & \beta \\
\beta^* & \alpha^*
\end{pmatrix} \in \SU(1,1).
\ee
Note that $\SU(1,1)\cong\SL(2,\mathbb R)$, which is as expected
since the three generators $L_0$ and $L_{q,\pm}$ in Eq.~\eqref{Hdeform_SL2} generate the $\SL^{(q)}(2,\mathbb R)$ group. 
Depending on the types of driving Hamiltonians in \eqref{3Types},
the $\SU(1,1)$ matrices in Eq.\eqref{Zmob} have different 
expressions as follows: 

\begin{enumerate}
    \item 
 Elliptic ($c^{(2)}<0$):
\be
\label{EllipticMobius}
\left\{
\begin{split}
&\alpha= \cos{\left( \frac{\pi \calC t}{l} \right)} + i \frac{\sigma^0}{\calC} \sin{\left( \frac{\pi \calC  t}{l} \right)},\\
&\beta= i \frac{\sigma^+ + i\sigma^-}{\calC} \sin{\left( \frac{\pi \calC  t}{l} \right)}.
\end{split}
\right.
\ee
\item Parabolic ($c^{(2)}=0$):
\be
\label{ParabolicMobius}
\left\{
\begin{split}
    &\alpha=1+ i  \frac{\sigma^0 \pi t}{l},\\
&\beta=i\frac{(\sigma^+ + i\sigma^-)\pi t }{l}.
\end{split}
\right.
\ee
\item Hyperbolic ($c^{(2)}>0$):
\be
\label{HyperbolicMobius}
\left\{
\begin{split}
&\alpha = \cosh{\left( \frac{\pi \calC  t}{l} \right)} + i\frac{\sigma^0}{\calC} \sinh{\left( \frac{\pi \calC t}{l} \right)},\\
&\beta=  i \frac{\sigma^+ + i\sigma^-}{\calC} \sinh{\left( \frac{\pi \calC  t}{l} \right)}.
\end{split}
\right.
\ee
\end{enumerate}
In all these three cases, we have defined the real number:
\be\label{Def_C}
\calC:= \sqrt{\big| -(\sigma^0)^2 + (\sigma^+)^2 + (\sigma^-)^2\big|}.
\ee
One can find that for different types of Hamiltonians, the 
corresponding M\"obius transformations are qualitatively different.
It is well known that there are in total three types of $\SU(1,1)$
matrices (See, e.g., Ref.~\onlinecite{simon2005orthogonal}) depending
on the value of their traces: For $|\text{Tr}(M)|<2$, $=2$, and $>2$, the corresponding $\SU(1,1)$ matrices are called elliptic, parabolic,
and hyperbolic matrices, respectively.
It is straightforward to check that for a general time duration $t$, where $t>0$,
the M\"obius transformations in Eqs.~\eqref{EllipticMobius}, 
\eqref{ParabolicMobius}, and \eqref{HyperbolicMobius}
are elliptic, parabolic, and hyperbolic, respectively.
For this reason, we denote the corresponding Hamiltonian types 
in \eqref{3Types}.

\bigskip

Now let us consider an $\SL_2$ deformed Floquet CFT with $N$ driving Hamiltonians.
Then the operator evolution after each period 
is determined by $z_N=\Pi_N\cdot z$ (and similarly for the 
anti-holomorphic part), where 
\be\label{Pi_N}
\Pi_N=
M_1\cdots M_N
=:
\left(
\begin{array}{cccc}
\alpha_N &\beta_N\\
\beta_N^* &\alpha_N^*
\end{array}
\right)\in\SU(1,1).
\ee
The operator evolution after $m$ periods of driving is determined 
by $\Pi_N$ as 
\be
\label{zn}
z_n=(\Pi_N)^m\cdot z,\quad n=mN.
\ee
Then the phase diagram of the Floquet CFT is determined by
$|\text{Tr}(\Pi_N)|$ as follows:~\cite{wen2018floquet,QuasiPeriodic}
\be\label{Pi_n_phase}
\left\{
\begin{split}
  &|\text{Tr}(\Pi_N)|<2, \quad \text{non-heating phase},\\ 
  &|\text{Tr}(\Pi_N)|=2, \quad \text{phase transition},\\ 
  &|\text{Tr}(\Pi_N)|>2, \quad \text{heating phase}.
\end{split}
\right.
\ee
As studied in Ref.~\onlinecite{wen2018floquet}, the value of $|\text{Tr}(\Pi_N)|$
in \eqref{Pi_n_phase} determines the trajectories of operator evolutions.
For $|\text{Tr}(\Pi_N)|<2$, the operator will rotate along the circle all the way;
for $|\text{Tr}(\Pi_N)|=2$, the operator will approach a fixed point polynomially fast
in time; for $|\text{Tr}(\Pi_N)|>2$, the operator will approach a fixed point exponentially
fast in time. As we will see in the next subsection,
the entanglement/energy-momentum density evolution will exhibit qualitatively different
features in these different regimes.

\subsection{Entanglement and energy-momentum evolution}
\label{Sec:EE_Energy}

Once we know the operator evolution in \eqref{Zmob}, 
we can study the time evolution of correlation functions.
The `order parameters' we use to distinguish different dynamical
phases are the entanglement entropy and energy-momentum evolutions,
which can be viewed as the correlation functions of twist operators
and energy-momentum tensor, respectively.
Some of related details can also be found in Refs.~\onlinecite{Wen:2018vux,wen2018floquet,fan2019emergent,QuasiPeriodic}.

Let us consider the time evolution of entanglement entropy first.
With the twist-operator approach,~\cite{Calabrese:2004eu,Calabrese_2009}
 one can find the $\alpha$-th Renyi entropy
of the subsystem $A=(x_1,\,x_2)$ as
\be
S^{(\alpha)}_A=\frac{1}{1-\alpha}\log\langle \Psi_n|\mathcal{T}(w_1,\bar{w}_1)
\bar{\mathcal{T}}(w_2,\bar{w}_2)|\Psi_n\rangle,
\ee
where $|\Psi_n\rangle$ is the wavefunction in \eqref{DiscreteDriving}
by choosing the initial state $|\Psi_0\rangle$ as the ground state of the 
uniform Hamiltonian $H_0$ in \eqref{H0},
and $\mathcal{T}$ ($\bar{\mathcal{T}}$) are twist operators 
with conformal dimensions $h=\bar{h}=\frac{c}{24}(\alpha-1/\alpha)$.
As studied in the previous subsection, to evaluate the correlation function 
of the twist operators, we first map the $w$-cylinder to the $q$-sheet
Riemann surface $z$ by $z=e^{i \frac{2\pi q w}{L}}$, where the evolution of $z_1 (\bar{z}_1)$
$z_2 (\bar{z}_2)$ are governed by M\"obius transformations in 
Eq.~\eqref{zn}.
Next, we map the $z$-Riemann surface to a complex plane $\zeta$ by $\zeta=z^{1/q}$.
One can find that 
\be
\begin{split}
&\langle \Psi_n|\mathcal{T}(w_1,\bar{w}_1)
\bar{\mathcal{T}}(w_2,\bar{w}_2)|\Psi_n\rangle\\
=&\prod_{i=1,2}
\left(\frac{\partial \zeta_i}{\partial w_i}\right)^h 
\prod_{i=1,2}
\left(\frac{\partial \bar{\zeta}_i}{\partial \bar{w}_{i}}\right)^{\bar{h}}
\langle 
\mathcal{T}(\zeta_1,\bar{\zeta}_1)
\bar{\mathcal{T}}(\zeta_2,\bar{\zeta}_2)
\rangle_{\zeta},
\end{split}
\ee
where we have $w_i=x_i+i\tau=x_i$.
For a general choice of the subsystem $A$, the expression of $S_A^{(\alpha)}$
is complicated.~\cite{wen2018floquet}
Here, for simplicity we choose the subsystem $A$ as a unit cell with
$(x_1,\,x_2)=(kl, \,(k+1)l)$ where $k\in\mathbb Z$. 
Then it is straightforward to find that
\be\label{EE_general1}
S_A(n)-S_A(0)=
\frac{c}{3}\Big(
\log\big|\alpha_n+\beta_n\big|+\log\big|\alpha'_n+\beta'_n\big|
\Big),
\ee
where $\alpha'_n$ ($\beta_n'$) corresponds to the driving effect in the 
anti-chiral parts and we have considered $S_A=\lim_{\alpha\to 1}S_A^{(\alpha)}$.
Here $\alpha_n$ and $\beta_n$ are the matrix elements in $(\Pi_N)^m$ in \eqref{zn}, i.e.,
\be
\label{AlphaBeta_n}
\begin{pmatrix}
\alpha_n &\beta_n\\
\beta_n^* &\alpha_n^*
\end{pmatrix}
=
\begin{pmatrix}
\alpha_N &\beta_N\\
\beta_N^* &\alpha_N^*
\end{pmatrix}^m, \quad n=mN.
\ee

As a remark, if one studies the entanglement entropy of $A$ by shifting a half unit cell, i.e., 
$A=[(k+\frac{1}{2})l, (k+\frac{3}{2})l]$ where $k\in\mathbb Z$, then one can find that
\be\label{EE_general2}
S_A(n)-S_A(0)=\frac{c}{3}
\Big(
\log|\alpha_n-\beta_n|+\log|\alpha'_n-\beta'_n|
\Big).
\ee
The difference between Eq.~\eqref{EE_general1} and Eq.~\eqref{EE_general2} 
reflects the fact that the system is driven in a non-uniform way.

Next, let us consider the energy-momentum density evolution.
With the operator evolution in \eqref{zn}, one has
\be\label{OP_evolution_T}
U^{\dag}\,T(z)\, U=\left(\frac{\partial z'}{\partial z}\right)^{2}
T\big(z'\big)+\frac{c}{12}\text{Sch}\{z',z\},
\ee
where $U=\big(U_N\cdots U_2\cdot U_1\big)^m$ [see Eq.~\eqref{DiscreteDriving}],
and the second term represents the Schwarzian derivative.
One can obtain the expectation value of the chiral energy-momentum tensor density as follows~
\cite{fan2019emergent}
\be\label{EnergyMomentum}
%\small
\frac{1}{2\pi}\langle T(x,n)\rangle=-\frac{q^2 \pi c}{12 L^2}+
\frac{\pi c}{12 L^2}\cdot(q^2-1)\cdot \frac{1}{|\alpha_n \cdot z+\beta_n|^4}
\ee
where $z=e^{\frac{2\pi i q x}{L}}$,
and $\alpha_n$ and $\beta_n$ are those defined in Eq.\eqref{AlphaBeta_n}.
By integrating over the energy-momentum density, one can obtain the 
total energy as
\be
\label{TotalEnergy}
\small
\begin{split}
E(n)=&\frac{1}{2\pi}\int_0^L\langle T(x,n)+ \bar{T}(x,n)\rangle dx\\
=&
-\frac{q^2 \pi c}{6 L}+
\frac{\pi c}{12 L}(q^2-1)\cdot(|\alpha_n|^2+|\beta_n|^2+|\alpha_n'|^2+|\beta_n'|^2)
\end{split}
\ee

Now let us comment on how different types of M\"obius transformations in 
Eqs.~\eqref{EllipticMobius}, \eqref{ParabolicMobius}, and \eqref{HyperbolicMobius}
result in different behaviors of entanglement/energy evolution. 
Based on the analysis in Refs.~\onlinecite{wen2018floquet,QuasiPeriodic}, the 
norm $|\alpha_n|$ ($|\beta_n|$) will grow exponentially/polynomically/oscillate
in time when the corresponding M\"obius transformations within one period is
hyperbolic/parabolic/elliptic.
Based on the expressions of entanglement and energy-momentum density
evolution in Eqs.~\eqref{EE_general1} and \eqref{TotalEnergy}, one can find that
in general the entanglement entropy will grow linearly/grow logarithmically/oscillate in time,
and the total energy will grow exponentially/grow polynomially/oscillate in time accordingly.
We will see an illustrating example later in Sec.~\ref{N1quench}.

\section{Classifying the $\SL_2$ deformed Floquet CFT}
\label{Sec:ClassifyFloquetCFT}

Now we come to the main section of this work: we study the  conditions for the 
existence of heating and non-heating phases in the phase diagrams when 
there are $N$ driving Hamiltonians.  
It is found that there are always heating phases in the phase diagram.
For the non-heating phases, we will give conditions for their existence 
in the phase diagram.
Our conditions are both sufficient and necessary for $N=1$ and $N=2$,
and are sufficient for $N>2$.
We will illustrate these conditions by considering $N=1$, $2$, and $3$, 
and then give the general results for arbitrary $N$.

\subsection{$N=1$}
\label{N1quench}

As a warm up, let us first consider the simplest case with $N=1$, 
i.e., there is only one driving Hamiltonian $H_1$. 
This case corresponds to a single quantum quench
rather than a Floquet CFT. Here we consider this simple
case to illustrate how different Hamiltonian types in~
\eqref{3Types} determine different behaviors of 
entanglement evolution.

It is noted that in Ref.~\onlinecite{Wen:2018vux}, 
the single quantum quench was studied for some specific 
Hamiltonians $H_1$ with $c^{(2)}=0$ and $c^{(2)}<0$, respectively.
It was found that the entanglement entropy grows 
logarithmically in time for $c^{(2)}=0$, and simply 
oscillates in time for $c^{(2)}<0$. Here we consider the more general Hamiltonians as described in Eqs.~\eqref{Hcfti}
and \eqref{fx_SL2}.

As discussed in the previous subsections, different types of 
Hamiltonians will determine different kinds of operator evolution
in Eqs.~\eqref{EllipticMobius}, \eqref{ParabolicMobius} and \eqref{HyperbolicMobius}, which further determine the
entanglement and energy-momentum density through Eqs.~\eqref{EE_general1} and \eqref{EnergyMomentum}.

Let us consider the entanglement entropy evolution for example.
Considering the subsystem $A=[kl, (k+1)l]$, where $k\in \mathbb Z$, 
it can be found that
in the long time driving limit (i.e., $\calC t/l\gg 1$ in the hyperbolic case,
and $t/l\gg 1$ in the parabolic case),  the entanglement entropy will grow linearly in time
for $c^{(2)}>0$ (hyperbolic case), grow logarithmically in time for $c^{(2)}=0$ (parabolic case),
and oscillate in time for $c^{(2)}<0$ (elliptic case) in the following way:
\be
\label{QuenchLongTime}
\small
S_A(t)-S_A(0)\simeq
\left\{
\begin{split}
&\frac{\pi c}{6}\cdot\frac{\calC t}{l}, \quad &c^{(2)}>0,\\
&\frac{c}{3}\log\frac{\pi t}{l},\quad &c^{(2)}=0,\\
&\frac{c}{6}\log\Big|
a+b\sin\big(
\frac{2\pi \mathcal{C} t }{l}+\phi
\big)
\Big|, &c^{(2)}<0,
\end{split}
\right.
\ee
where $\calC$ is defined in Eq.~\eqref{Def_C}. The real numbers 
$a$, $b$, and $\phi$ depend on the parameters $(\sigma^0,\sigma^+,\sigma^-)$. 
One can refer to Appendix.~\ref{Appendix:SingleQuench} for a complete expression
of the entanglement entropy evolution with arbitrary time duration $t>0$.

In short, the non-equilibrium dynamics for $N=1$ is only determined 
by the Hamiltonian types in \eqref{3Types}.

\subsection{$N=2$}
\label{Sec:N=2}

\begin{figure}[t]
\centering
\includegraphics[width = 3.4in]{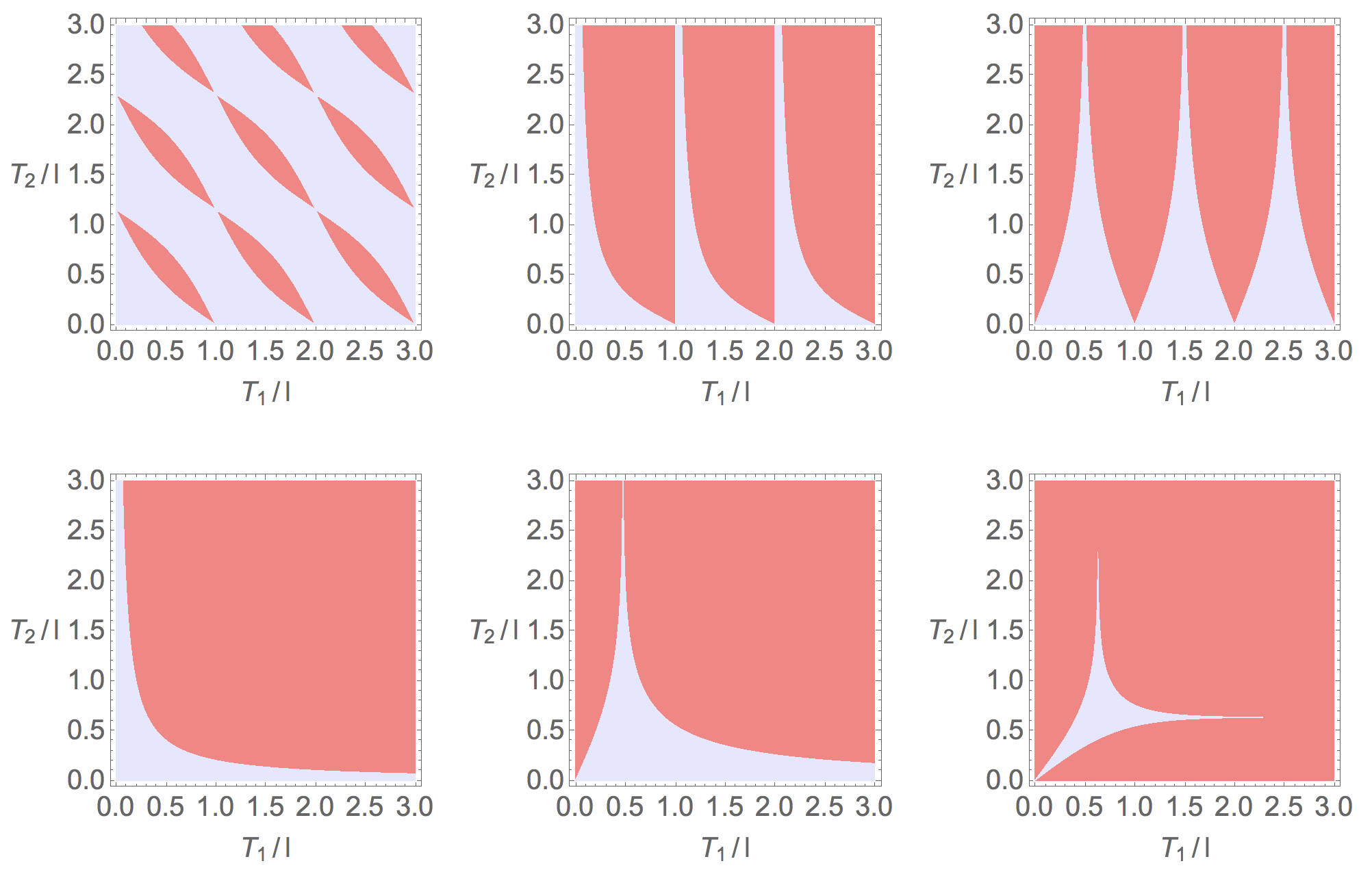}
\caption{
Phase diagrams of a Floquet CFT
with the both non-heating (in blue) and 
heating (in red) phases for
the six kinds of pairings with $N=2$ in Table.\ref{classify}.
The parameters are (from left to right, and then top to bottom):
elliptic-elliptic with
$\b{\calC}_1=(1,0,0)$ and $\b{\calC}_2=(1,0.5,0)$;
elliptic-parabolic with
$\b{\calC}_1=(1,0,0)$ and $\b{\calC}_2=(1,1,0)$;
elliptic-hyperbolic with
$\b{\calC}_1=(1,0,0)$ and $\b{\calC}_2=(0,0.4,0)$;
parabolic-parabolic with 
$\b{\calC}_1=(1,1,0)$ and $\b{\calC}_2=(1,0,1)$;
parabolic-hyperbolic with
$\b{\calC}_1=(1,1,0)$ and $\b{\calC}_2=(1,0.6,1)$;
hyperbolic-hyperbolic with
$\b{\calC}_1=(1,1.4,0)$ and $\b{\calC}_2=(1,0,1.4)$.
} 
\label{PhaseDiagram}
\end{figure}

Now we consider the properties of the phase diagram in the case of
$N=2$, i.e., there are two non-commuting Hamiltonians.
Since there are three possibilities of Hamiltonian types for 
$H_1$ ($H_2$), we have in total six different unordered pairings of $H_1$
and $H_2$ (See Table~\ref{classify}).

For later use, let us define the Casimir vector that characterizes the 
$\SL_2$ deformed Hamiltonian in \eqref{fx_SL2}:
\be\label{CasimirVector}
\b{\calC}_i=(\sigma^0_i, \sigma^+_i,\sigma^-_i),
\ee
as well as the product of two Casimir vectors
\be
\label{CasimirProduct}
\b{\calC}_i\cdot \b{\calC}_j:=-\sigma^0_i\sigma^0_j+\sigma^+_i \sigma^+_j
+\sigma^-_i\sigma^-_j.
\ee

Our strategy to determine the phase diagram
in the parameter space spanned by $\{(T_1/l, T_2/l)|\,0<T_1/l,
T_2/l <\infty\}$
can be briefly summarized as follows.
The effect of driving with Hamtiltonian $H_1$ ($H_2$)
for a time duration $T_1$ ($T_2$) are represented by a $\SU(1,1)$ matrix $M_1$ ($M_2$). 
Depending on the types of $H_1$ ($H_2$), the $\SU(1,1)$ matrix
$M_1$ ($M_2$) takes the form in one of Eqs.~\eqref{EllipticMobius},
\eqref{ParabolicMobius}, and \eqref{HyperbolicMobius}.
Then the phase diagram is 
determined by $|\text{Tr}(M_1\cdot M_2)|=|\text{Tr}(M_2\cdot M_1)|$ based on 
Eq.~\eqref{Pi_n_phase}.

Fig.~\ref{PhaseDiagram} is a sample plot of the phase diagrams 
for the six different pairings of $H_1$ and $H_2$ in Table~\ref{classify}.
The parameters in $\b{\calC}_1$ and $\b{\calC}_2$ are chosen such that 
there are both heating and non-heating phases in the phase diagram.
For arbitrary choices of $\b{\calC}_1$ and $\b{\calC}_2$, the heating phases
are generic, but the non-heating phases may be absent.
In Table~\ref{classify}, we give the sufficient and necessary conditions for the 
existence of non-heating phases in the phase diagram. The details of derivations of these
conditions can be found in Appendix~\ref{Appendix:N=2}.
\begin{table}[h]
\small
 \centering
\begin{tabular}{|l|c|c|c|}\hline
\diagbox[width=5em]{$\,\, H_1$}{$H_2\,\,$}&
  Elliptic & Parabolic & Hyperbolic \\ \hline
\,\, Elliptic & $\surd$  & $\surd$  &$\surd$  \\ [0.4ex]\hline
\, Parabolic & $\surd$  & $\b{\calC}_1 \cdot \b{\calC}_2 < 0$ &$\b{\calC}_1 \cdot \b{\calC}_2 < 0$ \\[0.4ex] \hline
Hyperbolic& $\surd$  & $\b{\calC}_1 \cdot \b{\calC}_2 < 0$&$1+\frac{\b{\calC}_1 \cdot \b{\calC}_2}{\calC_1\calC_2} < 0$ \\[0.6ex]
 \hline
\end{tabular}
\caption{Sufficient and necessary conditions for the existence of 
non-heating phases with $N=2$.
``$\surd$'' means the non-heating phases exist for arbitrary choices of
non-commuting Hamiltonians $H_1$ and $H_2$ of the corresponding types.
}\label{classify}
\end{table}

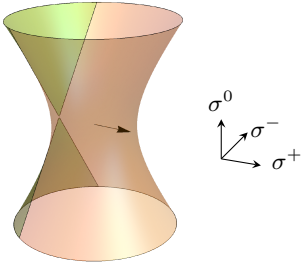
\begin{figure}[t]
\centering
\begin{tikzpicture}
\node[inner sep=0pt] (russell) at (0pt,0pt)
    {\includegraphics[width=.15\textwidth]{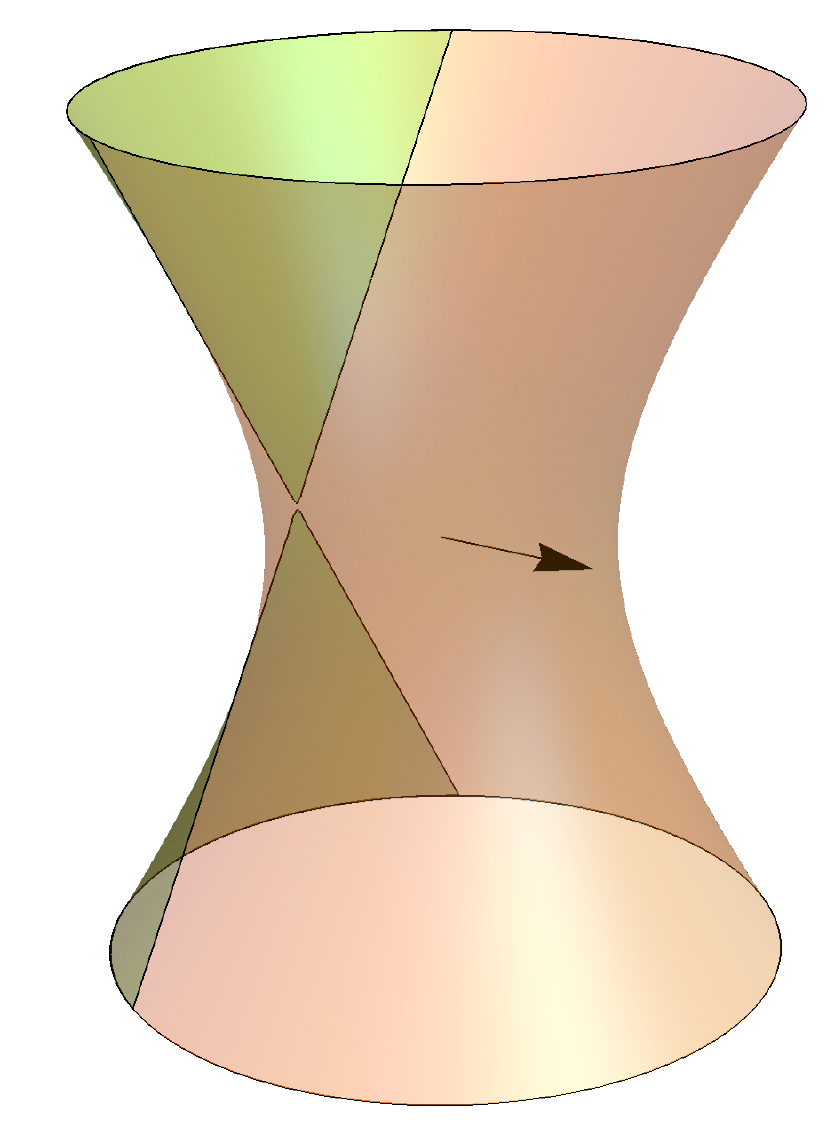}};
          \draw [>=stealth,->] (50pt, -10pt)--(65pt,-12.5pt);
          \draw [>=stealth,->] (50pt, -10pt)--(50pt,5pt);
           \draw [>=stealth,->] (50pt, -10pt)--(60pt,0pt);
                     \node at (75pt, -10pt){$\sigma^+$};
                     \node at (50pt, 12pt){$\sigma^0$};
                     \node at (67pt, 2pt){$\sigma^-$};
\end{tikzpicture}
\caption{$\frac{\b{\calC}_1}{\calC_1}=
(\sigma_1^0,\sigma_1^+,\sigma_1^-)=(0,1,0)$ 
is fixed (the vector in black).
The normalized vectors 
$\frac{\b{\calC}_2}{\calC_2}$ that satisfy the condition
in Eq.\eqref{HH_condition} 
are in the region in green.
} 
\label{Hyperbolic2ConditionFig}
\end{figure}

Let us give several remarks on how to obtain the conditions in Table~\ref{classify}. 
If at least one of the driving Hamiltonians is elliptic, then there must exist 
non-heating phases in the phase diagram. 
This can be straightforwardly understood as follows. 
Denoting the elliptic Hamiltonian as $H_1$, then the limit
$T_1/l\neq 0$ and $T_2/l=0$ ($j\neq i$) corresponds to a single
quench problem with $N=1$. In this case we have $|\text{Tr}(M_1\cdot M_2)|<2$.
Now we turn on $T_2/l$. As long as $T_2/l$ is small enough, we still have 
$|\text{Tr}(M_1\cdot M_2)|<2$, i.e., the Floquet CFT is in a non-heating phase.
Essentially, this condition is a quasi-($N-1$) condition, since only $(N-1)$ driving
Hamiltonians dominate while the left Hamiltonian plays little role.
The five conditions labeled by `$\surd$' in Table~\ref{classify} are all
quasi-$(N-1)$ conditions, with $N=2$. The left four conditions are 
intrinsic-$N$ conditions, which can be expressed by one condition:
\be\label{eta2}
\wideboxed{
\eta_2<0.
}
\ee
Here the indicator $\eta_{N=2}$ is constructed as follows.
For each driving Hamiltonian $H_j$, we arrange a matrix 
$P_j$ in the following way. If $H_j$ is parabolic, then 
\be
\label{PjParabolic}
P_j=\left(\begin{array}{cccc}
i\,\sigma_j^0 &i\,(\sigma_j^++i\sigma_j^-)\\
-i\,(\sigma_j^+-i\sigma_j^-) &-i\,\sigma_j^0
\end{array}
\right).
\ee
If $H_j$ is hyperbolic, then the corresponding matrix is
\be
\label{PjHyperbolic}
P_j=\left(\begin{array}{cccc}
1+i\,\frac{\sigma_j^0}{\calC_j} &i\,\frac{(\sigma_j^++i\sigma_j^-)}{\calC_j}\\
-i\,\frac{(\sigma_j^+-i\sigma_j^-)}{\calC_j} &1-i\,\frac{\sigma_j^0}{\calC_j}
\end{array}
\right).
\ee
Here $P_j$ are obtained based on the $\SU(1,1)$
matrix in \eqref{Zmob} as follows. For parabolic $H_j$, one has
\be
P_j=\lim_{t/l\to\infty}\left[
\left(\frac{\pi t}{l}\right)^{-1} M
\right],
\ee
with $M$ in \eqref{ParabolicMobius}. For hyperbolic $H_j$, one has
\be
P_j=\lim_{\calC t/l\to\infty}\left[
\left(\cosh\frac{\pi \calC t}{l}\right)^{-1} M
\right],
\ee
with $M$ in \eqref{HyperbolicMobius}.
Then the indicator is defined as
\be\label{eta2_expression}
\eta_2:=\text{Tr}(P_1\cdot P_2).
\ee
One can check explicitly that if at least one of the non-elliptic Hamiltonians is parabolic, then
the condition in \eqref{eta2} becomes
\be
\b{\calC}_1 \cdot \b{\calC}_2 < 0,
\ee
where the product of two Casimir vectors is defined in \eqref{CasimirProduct}.
If both of the two driving Hamiltonians are hyperbolic, then the condition in \eqref{eta2} becomes
\be
\label{HH_condition}
1+\frac{\b{\calC}_1 \cdot \b{\calC}_2}{\calC_1\calC_2} < 0,
\ee
The general principle of defining $\eta_2$ in \eqref{eta2_expression} 
can be straightforwardly understood as follows.
By writing the M\"obius transformations in Eqs.~\eqref{ParabolicMobius} and \eqref{HyperbolicMobius}
in terms of Pauli matrices, one can find that in the limit $\frac{T_i}{l}\to \infty$ ($i=1,2$)
the sum of coefficients of the leading terms in $\text{Tr}(M_1\cdot M_2)$ is nothing but $\eta_2$. If $\eta_2<0$,
then one has $\text{Tr}(M_1\cdot M_2)=-\infty$ in the limit $\frac{T_i}{l}\to \infty$ ($i=1,2$).
In the other limit $\frac{T_i}{l}\to 0$ ($i=1,2$), one can find that $\text{Tr}(M_1\cdot M_2)\to 2$.
Therefore, as we change the value of $\frac{T_i}{l}$ continuously, 
there must be a non-heating phase with $|\text{Tr}(M_1\cdot M_2)|<2$.

Now we give an illustration based on the hyperbolic-hyperbolic
driving. One can find details for other cases in Appendix~\ref{Appendix:N=2}. 
In the hyperbolic-hyperbolic driving, one can find that
\be\label{2HyperbolicTr}
\small
\begin{split}
&\text{Tr}(M_1\cdot M_2)=2\cosh\left(
\frac{\pi \mathcal{C}_1 T_1}{l}-\frac{\pi \mathcal{C}_2 T_2}{l}
\right)\\
&+2\left(1+
\frac{\b{\mathcal{C}}_1\cdot\b{\mathcal{C}}_2 }{\mathcal{C}_1\cdot \mathcal{C}_2}
\right)
\cdot \sinh\left(\frac{\pi \mathcal{C}_1 T_1}{l}\right)\cdot \sinh\left(\frac{\pi \mathcal{C}_2 T_2}{l}\right).
\end{split}
\ee
For $1+\frac{\b{\mathcal{C}}_1\cdot\b{\mathcal{C}}_2 }{\mathcal{C}_1\cdot \mathcal{C}_2}=0$, 
the Floquet CFT will stay at the phase transition 
(or critical phase) along the line
$
\frac{\calC_1 T_1}{l}=\frac{\calC_2 T_2}{l}.
$
Away from this critical line, the driven CFT will always be in the heating phase.
For $1+\frac{\b{\mathcal{C}}_1\cdot\b{\mathcal{C}}_1 }{\mathcal{C}_1\cdot \mathcal{C}_2}> 0$, one always has $\text{Tr}(M_1\cdot M_2)>2$ for arbitrary $0<T_1/l, \,T_2/l <\infty$, and 
therefore the system is always in the heating phase.
The non-heating phase can appear if and only if
\eqref{HH_condition} is satisfied,
which can be understood as follows. In the limit $T_1/l,\,T_2/l \to 0$, one has 
$\text{Tr}(M_1\cdot M_2)\to 2$. In the other limit $T_1/l,\,T_2/l \to \infty$, one has
$\text{Tr}(M_1\cdot M_2)\to -\infty$. By continuously changing $T_1/l$ and $T_2/l$,
 there must exist a region where $|\text{Tr}(M_1\cdot M_2)|< 2$, which
corresponds to the non-heating phase.
As an intuitive picture, by fixing $\frac{\b{\calC}_1}{\calC_1}=
(\sigma_1^0,\sigma_1^+,\sigma_1^-)=(0,1,0)$, the vector $\frac{\b{\calC}_2}{\calC_2}$
that satisfies~\eqref{HH_condition} is shown in Fig.~\ref{Hyperbolic2ConditionFig}.

\begin{figure}[t]
\centering
\includegraphics[width = 3.4in]{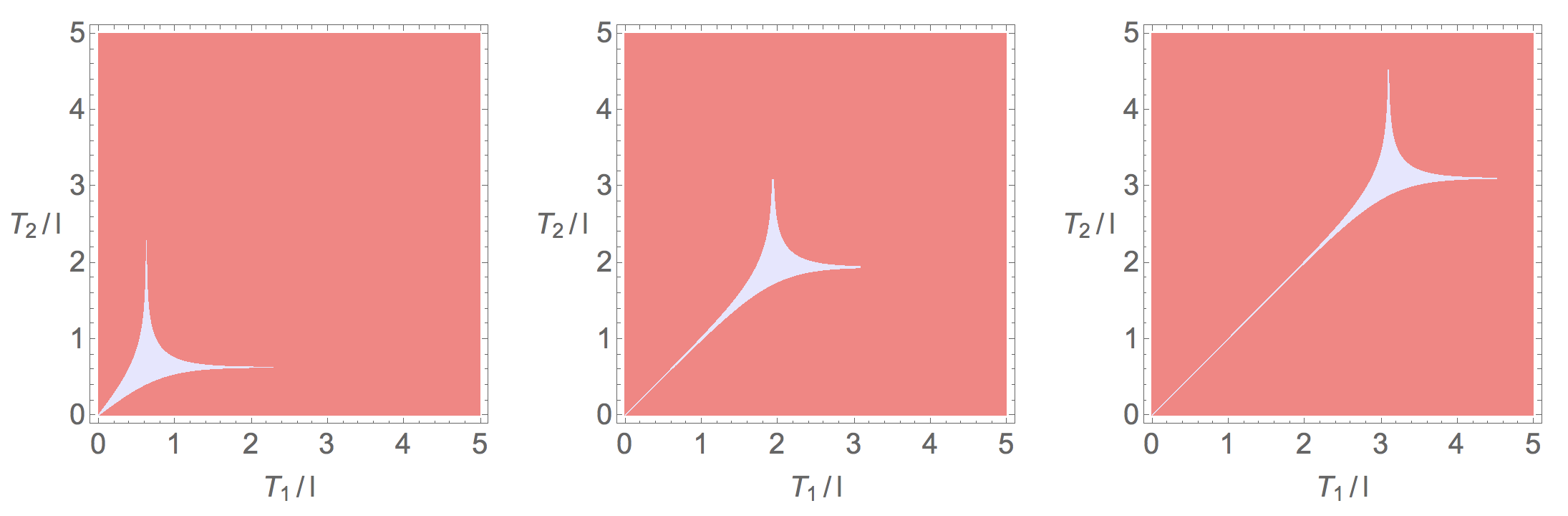}
\caption{
Phase diagram in a Floquet CFT with $N=2$ driving 
Hamiltonians,  both of which are of hyperbolic types.
The corresponding Casimir vectors are
$\b{\calC}_1=(1,\,a,\,0)$ and $\b{\calC}_2=(1,\,0,\, a)$, where we choose
$a=1.4$ (left), $1.41421$ (middle), and $1.41421356$ (right). 
The location of the non-heating phase in blue will 
move to infinity as we approach $a=\sqrt{2}$ from $a<\sqrt{2}$. 
For $a>\sqrt{2}$, the condition in \eqref{HH_condition} is violated, and
 the non-heating phase does not exist.
} 
\label{PhaseDiagramHyper}
\end{figure}

Before we leave this subsection, we hope to point out one interesting feature in 
the phase diagram of hyperbolic-hyberbolic driven Floquet CFT.
As we approach 
$1+\frac{\b{\mathcal{C}}_1\cdot\b{\mathcal{C}}_2 }{\mathcal{C}_1\cdot \mathcal{C}_2}=0$
from \eqref{HH_condition}, it is found that the non-heating phase does not vanish continuously.
What we observe is that the non-heating phase is composed of an island connected to three
lines  (see Fig.~\ref{PhaseDiagramHyper}). As we approach 
$1+\frac{\b{\mathcal{C}}_1\cdot\b{\mathcal{C}}_2 }{\mathcal{C}_1\cdot \mathcal{C}_2}=0$, 
this island of non-heating phase does not vanish but simply moves to the infinity (See 
Appendix~\ref{Appendix:N=2} for more details).

\subsection{$N=3$}
\label{Sec:N=3}
Now let us consider the case of $N=3$, i.e., there are three
driving Hamiltonians $H_1$, $H_2$, and $H_3$ in a driving period.
Similar to the previous subsection, we determine the phase 
diagram based on the value of $|\text{Tr}(M_1\cdot M_2\cdot M_3)|$
according to the criteria in \eqref{Pi_n_phase}.

One can find there are in general three layers of conditions to
ensure the existence of non-heating phases:

\begin{enumerate}

\item Quasi-$n$ (with $n=1$) condition. 
In this case, one simply needs to look at if there is an elliptic Hamiltonian
in the driving. If there is, then there must exist a non-heating phase in
the phase diagram. In particular, the non-heating phase can be realized 
when the elliptic Hamiltonian dominates in the driving.

\item Quasi-$n$ (with $n=2$) condition.
One needs to consider all possible choices of pairings $(i,j)$ in the driving.
If both Hamiltonians are non-elliptic,
then the quasi-$n$ ($n=2$) condition is
\be
\label{N3eta2}
\wideboxed{
\eta_{n=2}<0.
}
\ee
One can find that these conditions are nothing but those 
obtained in Table~\ref{classify}. In particular, the non-heating phases are realized
when $H_i$ and $H_j$ dominate in the driving.

\item Intrinsic-$N$ ($N=3$) condition. 
Now we need to consider all the three driving Hamiltonians together.
If all the three Hamiltonians are non-elliptic, then the intrinsic-$N$ ($N=3$) 
condition ensuring the existence of non-heating phases is
\be\label{N3eta3}
\wideboxed{
\eta_{N=3}<0,
}
\ee
where we have defined 
\be
\label{etaN3}
\eta_{N=3}:=\text{Tr}(P_1\cdot P_2\cdot P_3), 
\ee
with the matrices $P_j$ ($j=1,\,2,\,3$) of the form in Eqs.~\eqref{PjParabolic} or \eqref{PjHyperbolic}
depending on the types of Hamiltonian $H_j$.
\end{enumerate}

If at least one the above conditions is satisfied, then there must exist a non-heating phase.

The condition in \eqref{N3eta3} is obtained in a similar way to that in obtaining
\eqref{eta2}. That is, by tracking the behavior of $\text{Tr}(M_1\cdot M_2\cdot M_3)$
with $\frac{T_i}{l}$ ($i=1,\,2,\,3$) varied from $0^+$ to $\infty$, one can find that
the condition $\eta_{N=3}<0$ ensures the existence of non-heating phases. 

Let us consider examples on the expression of $\eta_{N=3}$ in \eqref{etaN3}
(One can find more examples in Appendix~\ref{Appendix:N3}).
If all the three Hamiltonians are parabolic, one has
\be
\eta_{N=3}=\b{\calC}_1 * \b{\calC}_2 * \b{\calC}_3,
\ee
where we have defined
\begin{align}
\label{C1C2C3}
\b{\calC}_1 * \b{\calC}_2 * \b{\calC}_3 :=& 
\sigma_1^{0} \sigma_2^{+} \sigma_3^{-} - \sigma_1^{0} \sigma_2^{-} \sigma_3^{+}
+ \sigma_1^{+} \sigma_2^{-} \sigma_3^{0} \\
&- \sigma_1^{+} \sigma_2^{0} \sigma_3^{-} + \sigma_1^{-} \sigma_2^{0} \sigma_3^{+}
- \sigma_1^{-} \sigma_2^{+} \sigma_3^{0}  \nonumber.
\end{align}
If all the three Hamiltonians are of hyperbolic types, then
$\eta$ can be expressed as
\be\label{etaHHH}
\eta_{N=3}=
1+\sum_{i<j} 
\frac{\b{\calC}_i \cdot \b{\calC}_j}{\calC_i \, \calC_j}
+ \frac{\b{\calC}_1 * \b{\calC}_2 * \b{\calC}_3}{\calC_1 \, \calC_2 \, \calC_3}.
\ee
Next, let us illustrate the intrinsic-$N$ ($N=3$) conditions in \eqref{N3eta3} 
explicitly with these two specific cases.

In the first illustrating case, all the three driving Hamiltonians are parabolic. Based on the M\"obius 
transformation in \eqref{ParabolicMobius}, one can obtain
\be
\label{TrPPP}
\small
\begin{split}
&\text{Tr}(M_1\cdot M_2\cdot M_3)=2\Big(1+\sum_{i<j} ^3
\, \b{\calC}_i \cdot \b{\calC}_j \, \frac{\pi T_i}{l}\cdot \frac{\pi T_j}{l}\\
&+\frac{\pi T_1}{l}\cdot \frac{\pi T_2}{l}\cdot \frac{\pi T_3}{l} \,
\b{\calC}_1 * \b{\calC}_2 *  \b{\calC}_3\Big).
\end{split}
\ee
For $\frac{\pi T_i}{l}\to 0^+$ ($i=1,\,2,\,3$), one has $\text{Tr}(M_1\cdot M_2\cdot M_3)\to 2$.
In the other limit  $\frac{\pi T_i}{l}\to \infty$, the last term in \eqref{TrPPP} will dominate.
If $\b{\calC}_1 * \b{\calC}_2 * \b{\calC}_3<0$, then we will have 
$\text{Tr}(M_1\cdot M_2\cdot M_3)\to-\infty$. That is, as we increase $\frac{\pi T_i}{l}$,
$\text{Tr}(M_1\cdot M_2\cdot M_3)$ changes from $2$ to $-\infty$ continuously. 
Apparently, there will be a non-heating phase (with $|\text{Tr}(M_1\cdot M_2\cdot M_3)|<2$)
in the parameter space.

In the second illustrating case, we consider three hyperbolic Hamiltonians. 
Based on the M\"obius transformations in \eqref{HyperbolicMobius}, one can find that
\be\label{Tr123}
\small
\begin{split}
&\text{Tr}(M_1 \cdot M_2 \cdot M_3) = 2\left[ \cosh{\frac{\pi  \calC_1 T_1}{l}} \cosh{\frac{\pi  \calC_2 T_2}{l}} \cosh{\frac{\pi\calC_3T_3}{l}} \right.  \\
&+\sum^3_{i<j;k\neq i,j} \frac{\b{\calC}_i \cdot \b{\calC}_j}{\calC_i \, \calC_j}
\sinh{\frac{\pi  \calC_i T_i}{l}} \sinh{\frac{\pi  \calC_j T_j}{l}} \cosh{\frac{\pi \calC_k T_k }{l}} \\
& \left. + \frac{\b{\calC}_1 * \b{\calC}_2 * \b{\calC}_3}{\calC_1 \, \calC_2 \, \calC_3} \sinh{\frac{\pi  \calC_1 T_1}{l}} \sinh{\frac{\pi  \calC_2 T_2}{l}} \sinh{\frac{\pi \calC_3 T_3 }{l}} \right].
\end{split}
\ee
As before, by choosing $\frac{\calC_iT_i}{l}\to 0^+$ ($i=1,\, 2,\,3$), one has
$\text{Tr}(M_1 \cdot M_2 \cdot M_3)\simeq 2$.
Next, by taking the limit $\frac{\calC_iT_i}{l}\to \infty$, 
one has $\cosh\frac{\pi\calC_i T_i}{l}\simeq \sinh\frac{\pi\calC_i T_i}{l}$, and therefore
$\text{Tr}(M_1 \cdot M_2 \cdot M_3)\simeq -2 \eta_{N=3}
(\cosh{\frac{\pi  \calC_1 T_1}{L}} \cosh{\frac{\pi  \calC_2 T_2}{l}} \cosh{\frac{\pi\calC_3T_3}{l}}) $,
where $\eta_{N=3}$ is defined in Eq.~\eqref{etaHHH}. 
If $\eta_{N=3}<0$, we will have 
$\text{Tr}(M_1 \cdot M_2 \cdot M_3)\to -\infty$. 
Therefore, as we tune $\frac{\calC_i T_i}{l}$ from $0^+$ to $\infty$,
the amplitude of $\text{Tr}(M_1\cdot M_2\cdot M_3)$ will change from $-2$ to $\infty$
continuously. Apparently, there will be a non-heating phase 
(with $|\text{Tr}(M_1\cdot M_2\cdot M_3)|<2$)
in the parameter space.

\begin{figure}
\centering
\includegraphics[width = 3.0in]{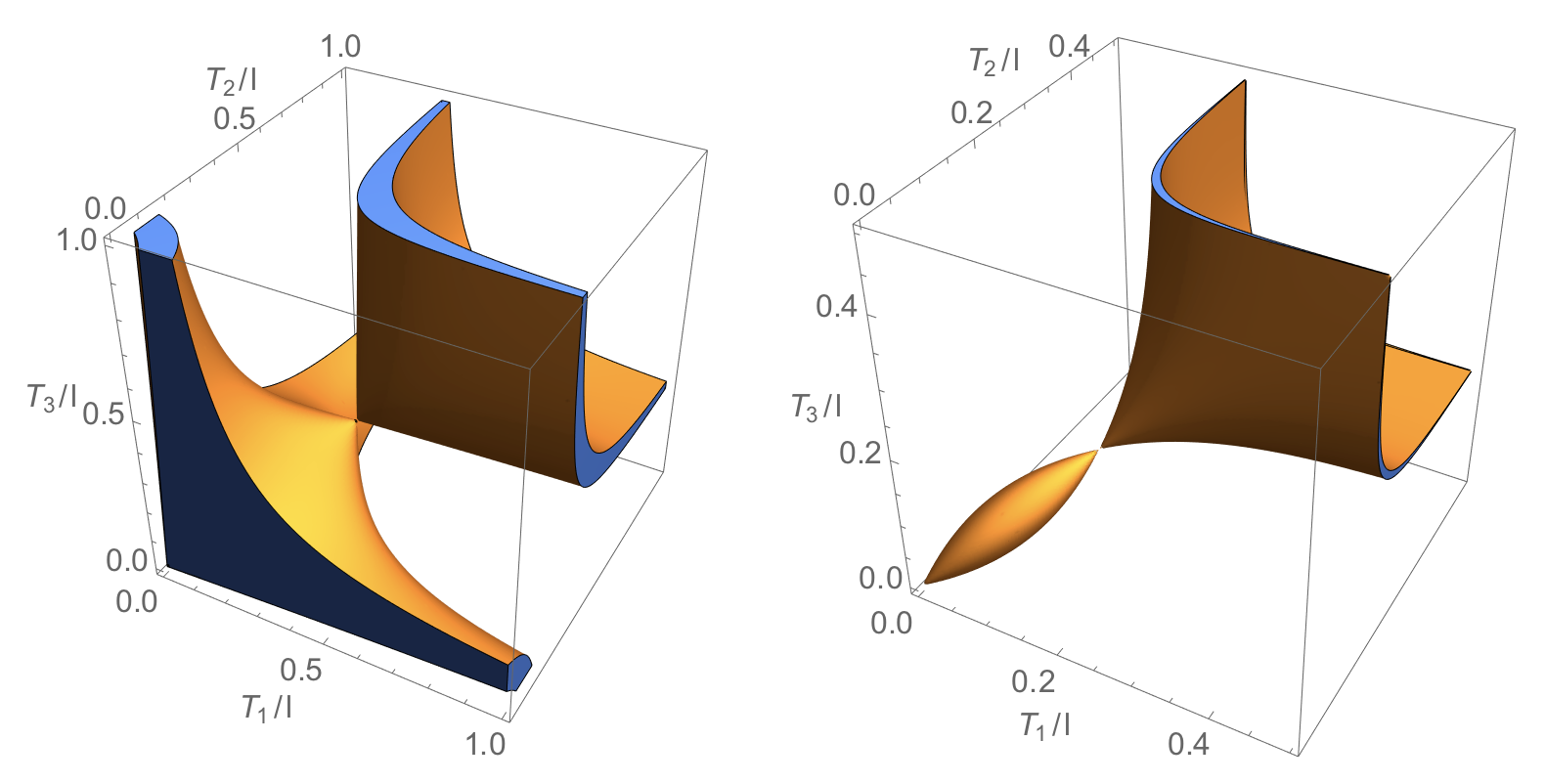}
\caption{
Non-heating phases in a Floquet CFT with $N=3$ driving 
Hamiltonians, all of which are of hyperbolic types.
We choose $\lambda=1.1$ (left) in \eqref{HHH} such that
only the condition $\eta_{n=2}<0$ is satisfied,
and $\lambda=3$ (right) such that only the condition
$\eta_{N=3}<0$ is satisfied.
The complemented regions are in the heating phase.
} 
\label{HHHphase}
\end{figure}

To have an intuitive picture of the phase diagram, 
we give a sample plot of the non-heating phases 
when the driving Hamiltonians are all hyperbolic.
We consider the Casimir vectors:
\be
\label{HHH}
\b{\calC}_i=(1, \lambda \cos\theta_i, \lambda \sin\theta_i),
\ee
where $\lambda>1$ such that the deformed Hamiltonian is always of hyperbolic type.
By choosing $\theta_1=0$, $\theta_2=\frac{2\pi}{3}$, and $\theta_3=\frac{4\pi}{3}$,
one can check explicitly that for $i,j\in 1, 2, 3$ and $i\neq j$, one always has
\be
\eta_2=
\frac{1}{2}-\frac{3}{2}\frac{1}{\lambda^2-1},
\ee
and 
\be
\eta_3=-\frac{1}{2}-\frac{9}{2}\cdot\frac{1}{\lambda^2-1}+\frac{3\sqrt{3}}{2}
\cdot\frac{\lambda^2}{(\lambda^2-1)^{3/2}}.
\ee
Then by solving the conditions in \eqref{N3eta2} and \eqref{N3eta3}, one can obtain
$\eta_2<0$ when $1<\lambda<2$, and $\eta_3<0$ when $\lambda>2$.
Typical plots of the phase diagrams for these two cases can be found in 
Fig.~\ref{HHHphase}.

\subsection{General $N$}
\label{Sec:General_N}

Based on the discussions in the previous subsections, now we are ready to give the conditions
for the existence of non-heating phases in an $\SL_2$ deformed Floquet CFT when there are 
$N$ driving Hamiltonians.

Let us denote the $N$ driving Hamiltonians as 
$\{H_1,\, H_2,\,\cdots, \, H_N\}$, which are arranged in time order of driving.
That is, we drive the CFT with $H_1$ for time duration $T_1$, 
$H_2$ for time duration $T_2$, and so on.
These driving Hamiltonians are characterized by Casimir vectors
$\{\b{\calC}_1,\, \b{\calC}_2,\,\cdots, \, \b{\calC}_N\}$ as defined in
\eqref{CasimirVector}.

The sufficient conditions for the existence of non-heating phases 
are composed of $N$ layers of conditions, which add constraints 
on the Casimir vectors $\{\b{\calC}_1,\, \b{\calC}_2,\,\cdots, \, \b{\calC}_N\}$. 
In layer $n$ ($1\le n \le N$),
we consider all possible choices of sets 
$\{H_{i_1},\cdots, H_{i_n}\}$, where the Hamiltonians are again arranged in the time
order of driving. Then the layer-$n$ conditions are:

\begin{enumerate}

\item If there is at least one elliptic Hamiltonian in the set 
$\{H_{i_1},\cdots, H_{i_n}\}$, then there is no constraint on the Casimir 
vectors $\{\b{\calC}_{i_1},\, \b{\calC}_{i_2},\,\cdots, \, \b{\calC}_{i_n}\}$.

\item If all the Hamiltonians in $\{H_{i_1},\cdots, H_{i_n}\}$ are non-elliptic
(either parabolic or hyperbolic), then the conditions ensuring the existence of
non-heating phases are
\be
\label{N_eta_n}
\wideboxed{
\exists \ \eta_{n}<0, \quad n=1,\cdots, N.
}
\ee
Here the indicator $\eta_n$ is defined as
\be
\label{eta_N}
\eta_n:=\text{Tr}(P_{i_1}\cdot P_{i_2}\cdots P_{i_n}),
\ee
with each matrix $P_j$ determined by the Casimir vectors 
$\b{\calC}_j=(\sigma_j^0,\sigma_j^+,\sigma_j^-)$ as follows. 
If the driving Hamiltonian $H_j$ is parabolic, then $P_j$ has the form 
\be
\label{PjParabolic2}
P_j=\left(\begin{array}{cccc}
i\,\sigma_j^0 &i\,(\sigma_j^++i\sigma_j^-)\\
-i\,(\sigma_j^+-i\sigma_j^-) &-i\,\sigma_j^0
\end{array}
\right).
\ee
If the driving Hamiltonian $H_j$ is hyperbolic, then $P_j$ has the form 
\be
\label{PjHyperbolic2}
P_j=\left(\begin{array}{cccc}
1+i\,\frac{\sigma_j^0}{\calC_j} &i\,\frac{(\sigma_j^++i\sigma_j^-)}{\calC_j}\\
-i\,\frac{(\sigma_j^+-i\sigma_j^-)}{\calC_j} &1-i\,\frac{\sigma_j^0}{\calC_j}
\end{array}
\right).
\ee

\end{enumerate}

By considering all possible $1\le n\le N$, there are in total $2^N-1$ conditions.
If at least one of these conditions is satisfied, then
there must exist non-heating phases in the phase diagram.

Now we give several remarks on the layer-$n$ conditions: 

-- The way to obtain condition \eqref{N_eta_n} is similar to
the examples considered in Sec.~\ref{Sec:N=2}
and \ref{Sec:N=3}. That is, $\eta_n$ corresponds to the sum of coefficients of the 
leading-order terms in $\text{Tr}(M_{i_1}\cdot M_{i_2}\cdots M_{i_n})$ in the limit $T_{i_k}/l\to\infty$
for all $i_k\in\{i_1, i_2,\cdots, i_n\}$. 
Then the condition in \eqref{N_eta_n} ensures that 
 $\text{Tr}(M_{i_1}\cdot M_{i_2}\cdots M_{i_n})$ will change from $2$ to $-\infty$
as we tune $T_{i_k}/l$ from $0$ to $\infty$ continuously. Then there must exist 
non-heating phases with $|\text{Tr}(M_{i_1}\cdot M_{i_2}\cdots M_{i_n})|<2$.

-- It is noted that there are in total $C_N^n=\frac{N!}{(N-n)!n!}$ conditions in layer-$n$ conditions.
In addition, as we have mentioned, if there exists at least one elliptic Hamiltonian in the chosen 
set $\{H_{i_1},\cdots, H_{i_n}\}$, 
 there is no constraint on the corresponding Casimir vectors 
$\{\b{\calC}_{i_1},\, \b{\calC}_{i_2},\,\cdots, \, \b{\calC}_{i_n}\}$. We hope to emphasize that
this does not mean there is no constraint on the layer-$n'$ condition when $n'<n$, because
it is totally possible there is no elliptic Hamiltonian in the subset 
$\{H_{i_1},\cdots, H_{i_{n'}}\}\subset \{H_{i_1},\cdots, H_{i_n}\}$.

-- In the specific case $n=1$, the layer-$n$ condition mentioned above is simply reduced to
the existence of an elliptic driving Hamiltonian.

\bigskip

Finally, let us comment on where to find these non-heating
phases in the $N$-dimensional parameter space spanned by
$\{\frac{T_1}{l}, \frac{T_2}{l},\cdots, \frac{T_N}{l}\}$. 
Suppose a certain layer-$n$
condition is satisfied,  then if there exists at least one
elliptic Hamiltonian (which we denote as $H_{i_m}$)
in the subset $\{H_{i_1},\cdots, H_{i_n}\}$, then the non-heating
phase can be obtained by taking all $T_i/l\to 0$ but with a
finite $T_{i_m}/l$; if all the $n$ Hamiltonians in 
$\{H_{i_1},\cdots, H_{i_n}\}$ are non-elliptic, then
the non-heating phases can be found in the $n$-dimensional
subspace spanned by $\{\frac{T_{i_1}}{l},
\frac{T_{i_2}}{l},\cdots,\frac{T_{i_n}}{l}\}$.
One can simply take
$\frac{T_{i_1}}{l}=\frac{T_{i_2}}{l}=\cdots =\frac{T_{i_n}}{l}:=
\frac{T^*}{l}$.
By increasing $\frac{T^*}{l}$ from $0$ to $\infty$ gradually,
one will necessarily find a non-heating phase. 
For example, one can refer to Fig.~\ref{PhaseDiagramHyper} for the 
case of $N=n=2$ and the right plot in Fig.~\ref{HHHphase}
for the case of $N=n=3$.

\section{Conclusion and Discussion}
\label{Sec:Conclusion}

In this paper, we have studied how the types of driving Hamtiltonians
affect the phase diagrams in an $\SL_2$ deformed Floquet CFT.
It is found that the heating phases are generic, but the non-heating
phases may be absent in the phase diagram.
We give the $N$-layer conditions  (with each layer of conditions
expressed in \eqref{N_eta_n} ) for the 
existence of non-heating phases in an $\SL_2$ deformed Floquet CFT
with $N$ driving Hamiltonians.
We showed that these conditions are sufficient and necessary for $N=2$.
For $N>2$, we only showed that these conditions are sufficient. 
In fact, for small $N$ with $N>2$, we have  scanned the
parameter space numerically and did not find any non-heating phases if 
the conditions in \eqref{N_eta_n} are violated.
We conjecture that our conditions in \eqref{N_eta_n} are also necessary
conditions. It is an interesting future problem to prove this conjecture.

Besides the types of driving Hamiltonians, the concrete 
driving sequences will also affect the phase diagram, 
such as quasi-periodic drivings and random drivings, 
as discussed in Refs.~\onlinecite{QuasiPeriodic,lapierre2020fine,RandomCFT}.
One simple form of the quasi-periodic drivings is the Fibonacci
quasi-periodic driving as studied in Refs.~\onlinecite{QuasiPeriodic,lapierre2020fine}
recently.
One way to obtain the phase diagram is to use a periodic driving
to approach the quasi-periodic driving, by taking larger and larger
driving period. Our conditions may be helpful to
understand how the phase diagram in a quasi-periodically driven
CFT depends on the types of driving Hamiltonians.
In the random drivings, the dependence of phase diagrams on 
the types of driving Hamiltonians will also exhibit very interesting
structures, as will be discussed in detail in Ref.~\onlinecite{RandomCFT}.

One interesting problem is to generalize the $\SL_2$ deformations
to the more general deformations, by choosing a general real
function $f(x)$ in \eqref{Hcfti},
where the underlying group structure
is the Virasoro group. 
Recently, in Ref.~\onlinecite{ageev2020deterministic},
the authors consider related problems in an opposite way. That is, 
one can start from a certain interesting conformal map on the complex $z$-plane, and 
map it back to the physical spacetime to find out the corresponding deformation
of the energy-momentum tensor. In general, this `mapping back' procedure
cannot be analytically done, and one  needs to perform numerical calculations.
In addition, we hope to emphasize that it is possible that the envelope function $f(x)$ 
[see Eq.~\eqref{Hcfti}] generated in this way
may be not a real function, which may result in non-Hermitian deformed Hamiltonians.
Nevertheless, one may use this method to search for interesting conformal
maps under which the driven CFT exhibit exotic features.

Another interesting problem is on the characterization 
of the Floquet CFTs. Previous works characterize the phase diagrams
based on either entanglement entropy or energy evolution.\cite{wen2018floquet,fan2019emergent,Zurich2019,
QuasiPeriodic,lapierre2020fine,RandomCFT} 
More detailed features of the time-dependent driven CFT
can be captured by the entanglement Hamiltonian (and its
spectrum), which
was recently used to study the non-equilibrium dynamics such as quantum quenches in (1+1)d CFTs 
both analytically\cite{Cardy_2016,Wen_2018} 
and numerically\cite{Di_Giulio_2019,Zhu2020}.
In the setup of Floquet CFTs with $\SL_2$ deformation, 
we expect that the entanglement Hamiltonians
in different phases of Floquet CFTs may be classified into
three types (see \eqref{3Types}) up to certain 
envelope functions. We will leave this problem to a future work.

\begin{acknowledgments}

We thank B.~Beri, R.~Fan, Y.~Gu, H.~Shapourian,
C. von Keyserlingk, A.~Ludwig, I.~Martin, S.~Ryu, T.~Tada, A.~Vishwanath,  A.~Wall and J.~Q.~Wu for useful discussions. 
X.~W. also thanks D.~Ageev, A.~Bagrov, and A.~Iliasov 
for communications on Ref.~\onlinecite{ageev2020deterministic}. 
B.~H.~is supported by ERC Starting Grant No.~678795 TopInSy. X.~W.~is supported by Gordon and Betty Moore Foundation's EPiQS initiative through Grant No.~GBMF4303 at MIT.
\end{acknowledgments}

\appendix

\section{$N=1$}
\label{Appendix:SingleQuench}

In this appendix, we give further detailed discussions on the time evolution of entanglement entropy 
after a single quantum quench.
Let us first consider the simple choice of subsystem $A=[n l, (n+1)l]$ where $l=L/n$.
Let us keep the anti-chiral part undeformed, and only focus
on the effect of deformation in the chiral part.
Then based on Eq.~\eqref{EE_general1} and Eqs.~\eqref{EllipticMobius}, \eqref{ParabolicMobius}, and
\eqref{HyperbolicMobius}, one can obtain
\begin{widetext}
\be
\small
S_A(t)-S_A(0)=
\left\{
\begin{split}
&\frac{c}{6}\log \left\{\Big[
\cosh\Big(\frac{\pi \calC t}{l}\Big)-\frac{\sigma^-}{\calC}\sinh\Big(\frac{\pi \calC t}{l}\Big)
\Big]^2+\Big[
\frac{\sigma^0+\sigma^+}{\calC}\cdot \sinh\Big(\frac{\pi \calC t}{l}\Big)
\Big]^2\right\}, \quad &c^{(2)}>0,\\
&\frac{c}{6}\log\left\{
\Big(1-\frac{\pi \sigma^- t}{l}\Big)^2+\Big(\frac{\pi(\sigma^0+\sigma^+)t}{l}\Big)^2
\right\},\quad &c^{(2)}=0,\\
&\frac{c}{6}\log\left\{
\Big[\cos\Big(\frac{\pi \calC t}{l}\Big)-\frac{\sigma^-}{\calC}\sin\Big(\frac{\pi\calC t}{l}\Big)\Big]^2
+\Big[
\frac{\sigma^0+\sigma^+}{\calC}\cdot \sin\Big(\frac{\pi \calC t}{l}\Big)
\Big]^2\right\},\quad &c^{(2)}<0.
\end{split}
\right.
\ee
\end{widetext}
For general choices of $(\sigma^0,\sigma^+,\sigma^-)$, one can find that 
in the long time driving limit $t/l\gg 1$, the entanglement entropy can be approximated
by the formulas in \eqref{QuenchLongTime} in the main text.
But there is one subtlety we hope to point out.
In the case of $c^{(2)}>0$, one can find that by choosing 
$\sigma^-\neq 0$ and $\sigma^0=\sigma^+=0$, one has
\be
S_A(t)-S_A(0)=-\frac{c}{6}\cdot\frac{\pi \calC t}{l}.
\ee
That is, the entanglement entropy \textit{decreases} linearly in time.
This phenomenon has been analyzed in Ref.~\onlinecite{QuasiPeriodic}.
The reason is that the entanglement cut and the energy-momentum density
peaks coincide with each other.
Intuitively, in the study of $S_A(t)$, one needs to introduce a UV cutoff at 
the entanglement cuts. Since the energy-momentum density peaks are also
located at the entanglement cuts, during the driving, the degree of freedom
that carries the entanglement between $A$ and its complement will accumulate
at the entanglement cut. Due to the UV cut-off, these degrees of freedom cannot
be detected by the entanglement entropy, which results in a decrease
in the entanglement entropy. To see the linear growth of the entanglement entropy
in this case,  one simply needs to shift the locations of entanglement cuts.
For example, by choosing $A=[(k+\frac{1}{2})l, (k+\frac{3}{2})l]$ where $k\in\mathbb Z$,
the entanglement entropy is expressed in \eqref{EE_general2}.
In this case, the entanglement cuts and energy-momentum density peaks do not
coincide with each other.
With the same choice of $\sigma^-\neq 0$ and $\sigma^0=\sigma^+=0$, one can 
find that the entanglement entropy grows linearly in time now.

\section{$N=2$}
\label{Appendix:N=2}

In this appendix, we give a derivation of the results in Table.~\ref{classify} in the main text.
That is, we consider six different pairings of $H_1$ and $H_2$ with
the Hamiltonian types in \eqref{3Types}: (i) elliptic-elliptic, (ii) elliptic-parabolic,
(iii) elliptic-hyperbolic, (iv) parabolic-parabolic, (v) parabolic-hyperbolic, and
(vi) hyperbolic-hyperbolic.
Some detailed features of the phase diagram will also be discussed.

\subsection{Features of phase diagram}

\textit{(i) Elliptic-elliptic}

If both the driving Hamiltonians are elliptic, then based on the 
M\"obius transformations in Eqs.~\eqref{EllipticMobius}, \eqref{ParabolicMobius}, 
\eqref{HyperbolicMobius}, we have
\be
\label{2ellipticTr}
\small
\begin{split}
&\text{Tr}(M_1\cdot M_2)=2\cos\left(\frac{\pi \mathcal{C}_1 T_1}{l}\right)\cdot \cos\left(\frac{\pi \mathcal{C}_2 T_2}{l}\right)\\
&\quad+\frac{2\, \b{\calC}_1 \cdot \b{\calC}_2 }{\mathcal{C}_1\cdot \mathcal{C}_2}\cdot \sin\left(\frac{\pi \mathcal{C}_1 T_1}{l}\right)\cdot \sin\left(\frac{\pi \mathcal{C}_2 T_2}{l}\right).
\end{split}
\ee
First, as proved in Appendix \ref{Sec:HeatingLine}, 
there always exists a heating phase 
along the lines in Eqs.~\eqref{HeatingLine1}
and \eqref{HeatingLine2} in the parameter space.

Second, let us prove there always exist non-heating phases
in the phase diagram.
It is noted that when both $H_1$ are $H_2$ are elliptic,
we always have $\Big|\frac{\b{\calC}_1 \cdot \b{\calC}_2}{\calC_1\calC_2}\Big|>1$, as discussed 
in Appendix \ref{Sec:HeatingLine}.
Let us consider the cases with 
$\frac{\b{\calC}_1 \cdot \b{\calC}_2}{\calC_1\calC_2} <-1$ and 
$\frac{\b{\calC}_1 \cdot \b{\calC}_2}{\calC_1\calC_2} >-1$ separately.

For $\frac{\b{\calC}_1 \cdot \b{\calC}_2}{\calC_1\calC_2} <-1$, it is convenient
to rewrite Eq.\eqref{2ellipticTr} as follows
\be
\small
\begin{split}
\text{Tr}(M_1\cdot M_2)=&2\cos\left(
\frac{\pi\calC_1T_1}{l}+\frac{\pi\calC_2T_2}{l}
\right)\\
&+2 \left(\frac{\b{\calC}_1 \cdot \b{\calC}_2}{\calC_1\calC_2} +1\right)\cdot 
\sin\left(\frac{\pi \mathcal{C}_1 T_1}{l}\right)\cdot \sin\left(\frac{\pi \mathcal{C}_2 T_2}{l}\right).
\end{split}
\ee
One can find that for $\frac{\calC_1T_1}{l}\simeq 0^+$
and $\frac{\calC_2 T_2}{l}\simeq 0^+$, one has 
$0<\text{Tr}(M_1\cdot M_2) <2$, and therefore the 
system is in a non-heating phase.

Similarly, for $\frac{\b{\calC}_1 \cdot \b{\calC}_2}{\calC_1\calC_2} >1$,
one can rewrite Eq.\eqref{2ellipticTr} as
\be
\small
\begin{split}
\text{Tr}(M_1\cdot M_2)=&2\cos\left(
\frac{\pi\calC_1T_1}{l}-\frac{\pi\calC_2T_2}{l}
\right)\\
&+2 \left(\frac{\b{\calC}_1 \cdot \b{\calC}_2}{\calC_1\calC_2} -1\right)\cdot 
\sin\left(\frac{\pi \mathcal{C}_1 T_1}{l}\right)\cdot \sin\left(\frac{\pi \mathcal{C}_2 T_2}{l}\right).
\end{split}
\ee
For $\frac{\calC_1T_1}{l}\simeq 0^+$
and $\frac{\calC_2 T_2}{l}\simeq 1-0^+$,
or $\frac{\calC_1T_1}{l}\simeq 1-0^+$
and $\frac{\calC_2 T_2}{l}\simeq 0^+$,
we have 
$-2<\text{Tr}(M_1\cdot M_2) <0$, and therefore the 
system is in a non-heating phase.

For the two regions corresponding to non-heating phases 
as discussed above, one can see Fig.~\ref{HeatLine}
for example.

In short, when the two non-commuting driving Hamiltonians 
are both elliptic, there are both heating and 
non-heating phases in the phase diagram.

\bigskip
\textit{(ii) Elliptic-parabolic}

Without loss of generality, we consider the case
that $H_1$ is elliptic, and $H_2$ is parabolic.
Then we have
\be
\small
\begin{split}
\text{Tr}(M_1\cdot M_2)=&2\cos\left(\frac{\pi \mathcal{C}_1 T_1}{l}\right)\\
&+\frac{2\,\b{\mathcal{C}}_1\cdot\b{\mathcal{C}}_2 }{\mathcal{C}_1}\cdot
\frac{\pi T_2}{l}\cdot  \sin\left(\frac{\pi \mathcal{C}_1 T_1}{l}\right),
\end{split}
\ee
where $\b{\calC}_1\cdot\b{\calC}_2\neq 0$.
For $\b{\calC}_1\cdot\b{\calC}_2< 0$, one can find that 
for $\frac{\pi \mathcal{C}_1 T_1}{l}\simeq 0^+$ 
and $T_2/l\simeq 0^+$, 
we always have $0<\text{Tr}(M_1\cdot M_2)<2$, and therefore
the system is in a non-heating phase.
On the other hand, for finite $\frac{\pi \mathcal{C}_1 T_1}{l}$,
as $T_2$ goes to infinity, we always have a heating phase.
For $\b{\calC}_1\cdot\b{\calC}_2> 0$,
one can find that for 
$\frac{\pi \mathcal{C}_1 T_1}{l}\simeq 1-0^+$ 
and $\frac{\pi \mathcal{C}_2 T_2}{l}\simeq 0^+$, 
we always have $0<\text{Tr}(M_1\cdot M_2)<2$, and therefore
the system is in a non-heating phase.

\bigskip
\textit{(iii) Elliptic-hyperbolic}

Without loss of generality, we consider the case
that $H_1$ is elliptic, and $H_2$ is hyperbolic.
Then we have
\be
\small
\begin{split}
\text{Tr}(M_1\cdot M_2)=&2\cos\left(\frac{\pi \mathcal{C}_1 T_1}{l}\right)\cdot \cosh\left(\frac{\pi \mathcal{C}_2 T_2}{l}\right)\\
&+\frac{2\, \b{\mathcal{C}}_1\cdot\b{\mathcal{C}}_1 }{\mathcal{C}_1\cdot \mathcal{C}_2}\cdot \sin\left(\frac{\pi \mathcal{C}_1 T_1}{l}\right)\cdot \sinh\left(\frac{\pi \mathcal{C}_2 T_2}{l}\right).
\end{split}
\label{TraceEH}
\ee
For $\b{\calC}_1\cdot\b{\calC}_2=0$, it is straightforward to check
that both the heating and non-heating phases can 
exist in the phase diagram. For example, the driven CFT is always
in the non-heating phase along the lines 
$\frac{\calC_1T_1}{l}=\frac{1}{2}+n$, where $n\in\mathbb Z$
(See Fig.~\ref{PhaseDiagram}).
If $\frac{\calC_1T_1}{l}\neq\frac{1}{2}+n$, the driven CFT
will be in the heating phase for large enough $\frac{T_2}{l}$.
For $\b{\calC}_1\cdot\b{\calC}_2\neq 0$, one can always
write Eq.~\eqref{TraceEH} in the form of 
$N\cos\left(\frac{\pi \mathcal{C}_1 T_1}{l}+\phi\right)$,
where $|N|$ increases exponentially with $\calC_2T_2/l$ for 
$\calC_2T_2/l\gg 1$. Then the system is in the non-heating phase
along the lines $\frac{\pi \mathcal{C}_1 T_1}{l}+\phi=(1/2+n)\pi$ where 
$n\in\mathbb Z$, and in the heating phase when 
$\frac{\pi \mathcal{C}_1 T_1}{l}+\phi\neq (1/2+n)\pi$
and $\frac{\calC_2 T_2}{l}$ is large enough.
Therefore, for arbitrary $\b{\calC}_1\cdot\b{\calC}_2$, 
we always have both heating and non-heating phases in the 
phase diagram.

\bigskip

\textit{(iv) Parabolic-parabolic}

If both the driving Hamiltonians are parabolic, we have
\be
\small
\begin{split}
\text{Tr}(M_1\cdot M_2)=&2+2\, \b{\mathcal{C}}_1\cdot \b{\mathcal{C}}_2
\cdot \frac{\pi T_1 }{l}
\cdot \frac{\pi T_2 }{l},
\end{split}
\ee
where $\b{\calC}_1\cdot\b{\calC}_2\neq 0$.
Since $T_1$, $T_2> 0$, one can find that there are both 
heating and non-heating phases if 
$\b{\calC}_1\cdot \b{\calC}_2<0$.
On the other hand, for $\b{\calC}_1\cdot \b{\calC}_2>0$,
one always has $\text{Tr}(M_1\cdot M_2)>2$ and there is 
only a heating phase.

\bigskip

\textit{(v) Parabolic-hyperbolic}

Without loss of generality, we consider the case
that $H_1$ is parabolic, and $H_2$ is hyperbolic.
Then we have
\be
\small
\begin{split}
\text{Tr}(M_1\cdot M_2)
=&2\cosh\left(\frac{\pi \mathcal{C}_2 T_2}{l}\right)\\
&+\frac{2\,\b{\mathcal{C}}_1\cdot\b{\mathcal{C}}_2 }{\mathcal{C}_2}\cdot 
\frac{\pi T_1}{l}\cdot
\sinh\left(\frac{\pi \mathcal{C}_2 T_2}{l}\right).
\end{split}
\ee
Recall that $T_1$, $T_2>0$, it is straightforward to check that
there are both heating and heating phases if 
$
\b{\mathcal{C}}_1\cdot \b{\mathcal{C}}_2< 0.
$
There is only a heating phase if 
$
\b{\mathcal{C}}_1\cdot \b{\mathcal{C}}_2\ge 0.
$

\bigskip
\textit{(vi) Hyperbolic-hyperbolic}

If both the driving Hamiltonians are hyperbolic, then we have
\be
\small
\label{HyperbolicTrace2M}
\begin{split}
\text{Tr}(M_1\cdot M_2)=&2\cosh\left(\frac{\pi \mathcal{C}_1 T_1}{l}\right)\cdot \cosh\left(\frac{\pi \mathcal{C}_2 T_2}{l}\right)\\
&+\frac{2\,\b{\mathcal{C}}_1\cdot\b{\mathcal{C}}_2 }{\mathcal{C}_1\cdot \mathcal{C}_2}\cdot \sinh\left(\frac{\pi \mathcal{C}_1 T_1}{l}\right)\cdot \sinh\left(\frac{\pi \mathcal{C}_2 T_2}{l}\right),
\end{split}
\ee
which may be rewritten as
\be\label{2HyperbolicTr}
\small
\begin{split}
\text{Tr}(M_1\cdot M_2)&=2\cosh\left(
\frac{\pi \mathcal{C}_1 T_1}{l}-\frac{\pi \mathcal{C}_2 T_2}{l}
\right)\\
&+2\left(1+
\frac{\b{\mathcal{C}}_1\cdot\b{\mathcal{C}}_2 }{\mathcal{C}_1\cdot \mathcal{C}_2}
\right)
\cdot \sinh\left(\frac{\pi \mathcal{C}_1 T_1}{l}\right)\cdot \sinh\left(\frac{\pi \mathcal{C}_2 T_2}{l}\right)
\end{split}
\ee
For $\frac{\b{\mathcal{C}}_1\cdot\b{\mathcal{C}}_2 }{\mathcal{C}_1\cdot \mathcal{C}_2}=-1$, the driven CFT
will stay at the phase transition (or critical phase) along the line
\be\label{LineParabolic}
\frac{\calC_1 T_1}{l}=\frac{\calC_2 T_2}{l}.
\ee
Away from this critical line, the driven CFT will always be in the heating phase.

For $\frac{\b{\mathcal{C}}_1\cdot\b{\mathcal{C}}_1 }{\mathcal{C}_1\cdot \mathcal{C}_2}> -1$, one always has $\text{Tr}(M_1\cdot M_2)>2$, and 
therefore the system is in the heating phase.

The non-heating phase can appear if and only if
\be
\frac{\b{\mathcal{C}}_1\cdot\b{\mathcal{C}}_1 }{\mathcal{C}_1\cdot \mathcal{C}_2}< -1.
\ee
It is interesting to check how the non-heating phases 
disappear as $\frac{\b{\mathcal{C}}_1\cdot\b{\mathcal{C}}_1 }{\mathcal{C}_1\cdot \mathcal{C}_2}$ approaches $-1$ 
from the side of $\frac{\b{\mathcal{C}}_1\cdot\b{\mathcal{C}}_1 }{\mathcal{C}_1\cdot \mathcal{C}_2}<-1$.
For convenience, we denote 
$\frac{\b{\mathcal{C}}_1\cdot\b{\mathcal{C}}_1 }{\mathcal{C}_1\cdot \mathcal{C}_2}=-1-\epsilon$, where
$\epsilon=0^+$. In this limit, Eq.~\eqref{2HyperbolicTr} can be rewritten as
\be
\small
\begin{split}
\text{Tr}(M_1\cdot M_2)&=2\cosh\left(
\frac{\pi \mathcal{C}_1 T_1}{l}-\frac{\pi \mathcal{C}_2 T_2}{l}
\right)\\
&-2\epsilon
\cdot \sinh\left(\frac{\pi \mathcal{C}_1 T_1}{l}\right)\cdot \sinh\left(\frac{\pi \mathcal{C}_2 T_2}{l}\right)
\end{split}
\ee
One can find that the non-heating phases are composed of three
lines connected by a tri-junction (or island),
as shown in Fig.~\ref{PhaseDiagramHyper} for example.
This can be understood as follows.
First, let us consider the line along that defined in
Eq.~\eqref{LineParabolic}. By requiring $\text{Tr}(M_1\cdot M_2)=0$, one can obtain
\be
\label{Hyperbolic2nonheating01}
\small
\frac{\calC_1T_1}{l}=\frac{\calC_2T_2}{l}
=\frac{1}{\pi}\text{arcsinh}\sqrt{\frac{1}{\epsilon}}.
\ee
For all $\frac{\calC_1T_1}{l}$, 
$\frac{\calC_2T_2}{l}<\frac{1}{\pi}\text{arcsinh}\sqrt{\frac{1}{\epsilon}}$ along the line in Eq.~\eqref{LineParabolic}, 
one has $0<\text{Tr}(M_1\cdot M_2)<2$ and therefore the 
driven CFT is in a non-heating phase.
It is noted that as $\epsilon\to 0^+$, Eq.~\eqref{Hyperbolic2nonheating01} can be simplified as follows
\be
\small
\frac{\calC_1T_1}{l}=\frac{\calC_1T_1}{l}\simeq 
\frac{1}{\pi}\log\frac{2}{\epsilon}.
\ee
The upper boundary of the non-heating phase along the 
line in \eqref{LineParabolic} can be obtained by 
considering $\text{Tr}(M_1\cdot M_2)=-2$, based on which 
one can obtain $\frac{\calC_1T_1}{l}=\frac{\calC_2T_2}{l}
=\frac{1}{\pi}\text{arcsinh}\sqrt{\frac{1}{\epsilon}}$.

There are several interesting features:
(i) The non-heating phases are composed of three lines connected by an island.
(ii) As we approach $\frac{\b{\mathcal{C}}_1 }{\mathcal{C}_1}\cdot\frac{\b{\mathcal{C}}_2 }{\mathcal{C}_2}=-1$ from the side of $\frac{\b{\mathcal{C}}_1 }{\mathcal{C}_1}\cdot\frac{\b{\mathcal{C}}_2 }{\mathcal{C}_2}<-1$, the island will move to the infinity.

As a summary of this appendix, for $N=2$ non-commuting driving 
Hamiltonians, the phase diagram in the parameter space spanned by
$T_1/l$ and $T_2/l$ depends on the Hamiltonian types of both
$H_1$ and $H_2$.
If at least one of the two Hamiltonians is elliptic, then there must
exist non-heating phases in the phase diagram.

\subsection{Heating line in the elliptic-elliptic driving}
\label{Sec:HeatingLine}

In this appendix, we show that for two arbitrary non-commuting 
driving Hamiltonians that are elliptic, there always exist heating 
lines (lines along which the system is in heating phases)
in the phase diagram.

For later use, let us first define the reflection matrix:
$M\in \SU(1,1)$ is called \textit{reflection} if $M^2=-\mathbb I$
and $\text{Tr}(M)=0$~ \cite{simon2005orthogonal}.

From the definition, a reflection matrix is always elliptic.
In addition, it can be proved that the product of two non-commuting
reflection matrices is always hyperbolic~\cite{simon2005orthogonal}.
In other words, if both $M_1$ and $M_2$ are reflection matrices, 
and they do not commute with each other, then we always have
$|\text{Tr}(M_1\cdot M_2)|>2$.

\begin{figure}
\centering
\begin{tikzpicture}
\node[inner sep=0pt] (russell) at (0pt,0pt)
    {\includegraphics[width=.13\textwidth]{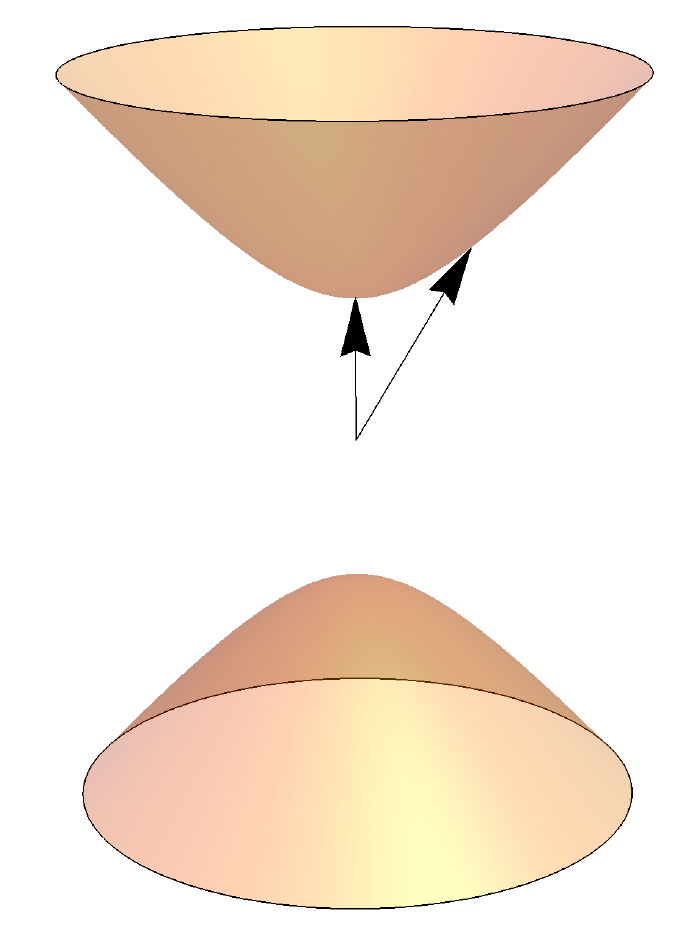}};
\node at (-10pt, 10pt){$\frac{\b{\calC}_1}{\calC_1}$};
\node at (20pt, 10pt){$\frac{\b{\calC}_2}{\calC_2}$};
\end{tikzpicture}
\caption{
Two vectors $\frac{\b{\calC}_1}{\calC_1}$ and 
$\frac{\b{\calC}_2}{\calC_2}$ corresponding to two 
non-commuting reflection matrices $M_1$ and $M_2$ 
in \eqref{2ReflectionProduct}.
} 
\label{2elliptic}
\end{figure}

Based on the above discussions, we can find that there are always
heating phases in the phase diagrams, if both the two non-commuting
driving Hamiltonians are of elliptic types.
The proof of this claim is as follows:

We consider two non-commuting drivings with $(H_1, T_1)$
and $(H_2, T_2)$, the corresponding M\"obius transformations
are expressed in Eq.~\eqref{EllipticMobius}.
One can find that by choosing 
\be\label{HeatingPoint}
T_j=\frac{l}{2\, \calC_j},
\ee
where $j=1,2$, one has
$\alpha_j=-i\frac{\sigma_j^0}{\calC_j}$,
$\beta_j=-i\frac{\sigma_j^++i\sigma_j^-}{\calC_j}$,
and 
\be\label{2ReflectionProduct}
\text{Tr}(M_1\cdot M_2)=\text{Tr}(M_2\cdot M_1)
=2\frac{\b{\calC}_1 \cdot \b{\calC}_2}{\calC_1\, \calC_2}.
\ee
For non-commuting elliptic Hamiltonians $H_1$ and $H_2$,
one can find that $|\text{Tr}(M_1\cdot M_2)|>2$.
This can be understood as follows.
One can always find a $\SU(1,1)$ matrix $U$, such that
$\text{Tr}(UM_1U^{-1}\cdot UM_2 U^{-1})=2\frac{\b{\calC}'_1 \cdot \b{\calC}'_2}{\calC'_1\calC'_2}$, where
the normalized vector $\frac{\b{\calC}'_1}{\calC_1'}$
is rotated to $(i,0,0)$ or $(-i,0,0)$.
Since $M_2\neq \pm M_1$, we have 
$\frac{\b{\calC}'_2}{\calC_2}=
(i\sigma^0_2\,',\sigma^+_2\,',\sigma^-_2\,')$,
with $-(\sigma^0_2\,')^2+(\sigma^+_2\,')^2+(\sigma^-_2\,')^2=-1$.
Since at least one of $\sigma^+_2\,'$ or $\sigma^-_2\,'$ is nonzero,
we always have $(\sigma^0_2\,')^2>1$ (See Fig.~\ref{2elliptic}). 

Therefore, 
$|\text{Tr}(M_1\cdot M_2)|=|2\sigma_2^0\,'|>2$,
or equivalently 
\be\label{C1C2elliptic}
\Big|\frac{\b{\calC}_1 \cdot \b{\calC}_2}{\calC_1\calC_2}\Big|>1.
\ee

In short, by choosing two 
non-commuting Hamtiltonians which are both elliptic,
we always have a heating phase at the point
$(T_1, T_2)=(\frac{l}{2\calC_1}, \frac{l}{2\calC_2})$.

Now we will show that there always exits a `heating line'
in the phase diagram if both driving Hamiltonians are elliptic.

\begin{figure}
\centering
\includegraphics[width = 2.5in]{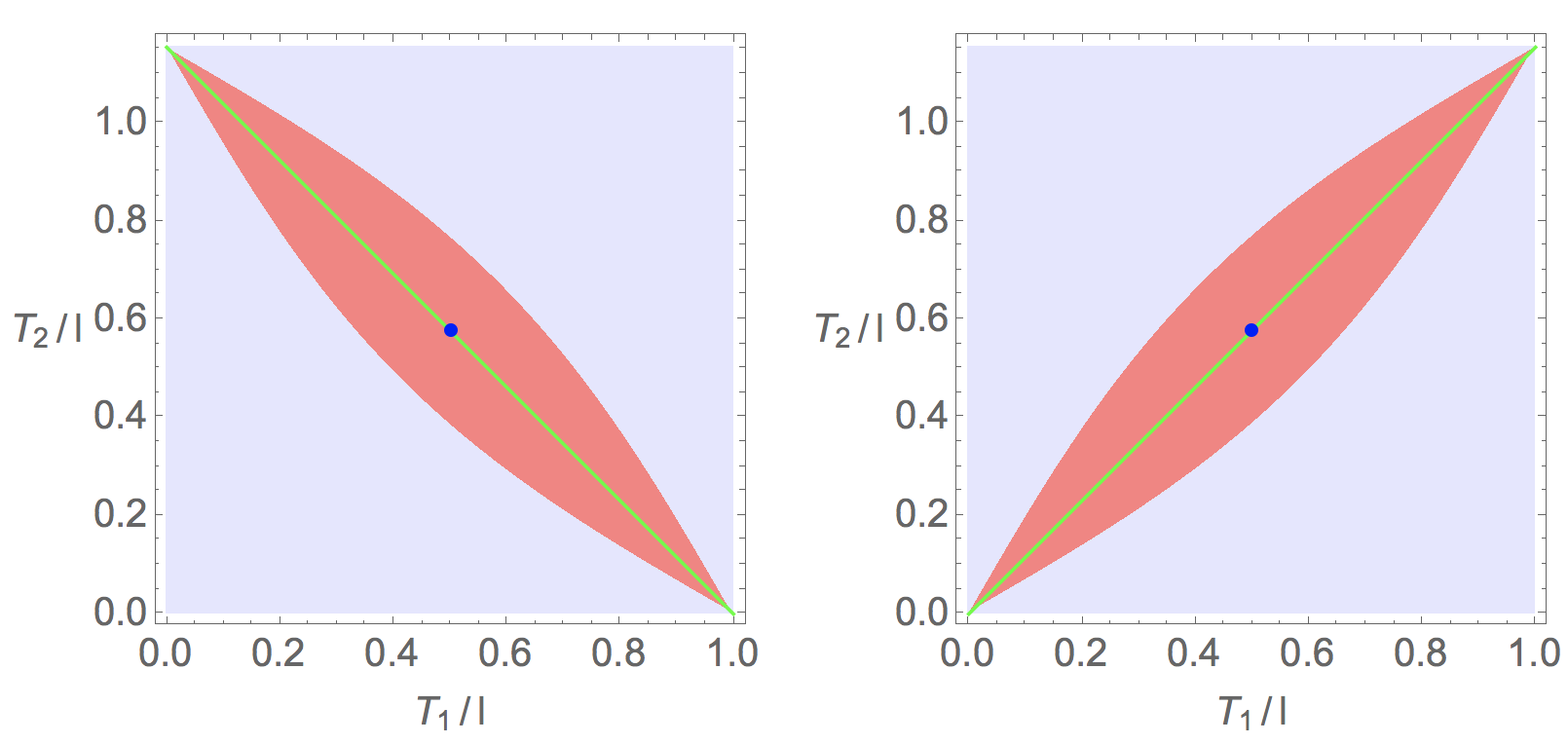}
\caption{
Heating line (in green) determined by Eqs.\eqref{HeatingLine1}
and \eqref{HeatingLine2} in the unit cell with 
$0<T_j\le l/\calC_j$ ($j=1,2$)
in the cases of
$\b{\calC}_1=(1,0,0)$ and $\b{\calC}_2=(1,0.5,0)$ (left)
and $\b{\calC}_1=(1,0,0)$ and $\b{\calC}_2=(-1,0.5,0)$ (right).
Here we choose $l=1$.
The blue dots represent the heating point 
as defined in Eq.\eqref{HeatingPoint}.
The regions in red (blue) corresponds to a
heating (non-heating) phase.
} 
\label{HeatLine}
\end{figure}

Now we only focus on a `unit cell' with 
$0<T_j\le l/\calC_j$ ($j=1,2$) in the phase diagram. 
The locations of \textit{heating line} depends on the sign of 
$\b{\calC}_1\cdot \b{\cal{C}}_2$ as follows:
\begin{enumerate}
    \item If $\b{\calC}_1\cdot \b{\cal{C}}_2<0$, the heating line 
    is determined by
    \be\label{HeatingLine1}
    \frac{T_1}{l_{1,\text{eff}}}+\frac{T_2}{l_{2,\text{eff}}}=1,\quad
    0<T_1<l_{1,\text{eff}},
    \ee
    where we have defined $l_{i,\text{eff}}=l/\calC_i$ in the elliptic case.
    \item If $\b{\calC}_1\cdot \b{\cal{C}}_2>0$, the heating line 
    is determined by
    \be\label{HeatingLine2}
    \frac{T_1}{l_{1,\text{eff}}}-\frac{T_2}{l_{2,\text{eff}}}=0,
    \quad
    0<T_1<l_{1,\text{eff}}.
    \ee
\end{enumerate}
Examples corresponding to these two cases can be found in Fig.~\ref{HeatLine}.
Now we give the proofs of these two claims as follows.

Let us consider $\b{\calC}_1\cdot \b{\cal{C}}_2<0$ first. Based on Eqs.~\eqref{HeatingLine1} and \eqref{2ellipticTr}
one can find that 
\be
\small
\text{Tr}(M_1\cdot M_2)=-2\cos^2\left(\frac{\pi T_1}{l_{1,\text{eff}}}\right)
+2\frac{\b{\calC}_1 \cdot \b{\calC}_2}{\calC_1\calC_2} 
\cdot \sin^2\left(\frac{\pi T_1}{l_{1,\text{eff}}}\right)
\ee
Since $\b{\calC}_1\cdot \b{\cal{C}}_2<0$, we have $\frac{\b{\calC}_1 \cdot \b{\calC}_2}{\calC_1\calC_2}<-1$ 
based on Eq.~\eqref{C1C2elliptic}.
Then one can find 
\be
\small
\text{Tr}(M_1\cdot M_2)=-2+2\left(\frac{\b{\calC}_1 \cdot \b{\calC}_2}{\calC_1\calC_2}+1\right)\cdot \sin^2\left(\frac{\pi T_1}{l_{1,\text{eff}}}\right)<-2.\nonumber
\ee
Therefore, we always have a heating phase along the line defined in 
Eq.~\eqref{HeatingLine1}.

Second, let us consider $\b{\calC}_1\cdot \b{\cal{C}}_2>0$. Based on Eqs.~\eqref{HeatingLine2} and \eqref{2ellipticTr}
one can find that 
\be
\small
\text{Tr}(M_1\cdot M_2)=2
+2\left(\frac{\b{\calC}_1 \cdot \b{\calC}_2}{\calC_1\calC_2}-1\right)\cdot \sin^2\left(\frac{\pi T_1}{l_{1,\text{eff}}}\right)>2,
\nonumber
\ee
where we have considered $\b{\calC}_1\cdot \b{\cal{C}}_2>0$ and 
therefore $\frac{\b{\calC}_1 \cdot \b{\calC}_2}{\calC_1\calC_2}>1$
based on Eq.~\eqref{C1C2elliptic}.

Till now, we have proved that the lines in 
Eq.~\eqref{HeatingLine1} for $\b{\calC}_1\cdot \b{\cal{C}}_2<0$
and those in Eq.~\eqref{HeatingLine2} for $\b{\calC}_1\cdot \b{\cal{C}}_2>0$
are always in the heating phase.

\section{$N=3$}
\label{Appendix:N3}

As discussed in Sec.~\ref{Sec:N=3}, when all the three driving Hamiltonians
are non-elliptic, the intrinsic-$N$ ($N=3$) condition is 
$\eta_N<0$ in \eqref{N3eta3}.
In the main text, we consider the cases with (i) three parabolic 
Hamiltonians and (ii) three hyperbolic Hamiltonians.
In this appendix, we consider the other two cases, i.e., 
(iii) one parabolic and two hyperbolic Hamiltonians and (iv) 
two parabolic and one hyperbolic Hamiltonians.

\textit{-- 1 parabolic and 2 hyperbolic Hamiltonians}

Now let us consider the case there are one parabolic and two hyperbolic driving Hamiltonians. 
Without loss of generality, let us choose $H_1$ is parabolic, and $H_2,\,H_3$
are hyperbolic.
Based on Eqs.~\eqref{ParabolicMobius} and \eqref{HyperbolicMobius}, one has
\be
\small
\begin{split}
&\text{Tr}(M_1\cdot M_2\cdot M_3)=2\cosh\left(\frac{\pi \calC_2 T_2}{l}\right)
\cosh\left(\frac{\pi \calC_3 T_3}{l}\right)\\
&+2\frac{\b{\calC}_1\cdot \b{\calC}_3}{\calC_3}\cdot \frac{\pi T_1}{l}\cdot
\cosh\left(\frac{\pi \calC_2 T_2}{l}\right)\sinh\left(\frac{\pi \calC_3 T_3}{l}\right)\\
&+2\frac{\b{\calC}_1\cdot \b{\calC}_2}{\calC_2}\cdot \frac{\pi T_1}{l}\cdot
\cosh\left(\frac{\pi \calC_3 T_3}{l}\right)\sinh\left(\frac{\pi \calC_2 T_2}{l}\right)\\
&+2\frac{\b{\calC}_2\cdot \b{\calC}_3}{\calC_2\,\calC_3}\cdot 
\sinh\left(\frac{\pi \calC_2 T_2}{l}\right)\sinh\left(\frac{\pi \calC_3 T_3}{l}\right)\\
&+2\frac{\b{\calC}_1*\b{\calC}_2*\b{\calC}_3}{\calC_2\,\calC_3}\cdot \frac{\pi T_1}{l}\cdot
\sinh\left(\frac{\pi \calC_2 T_2}{l}\right)\sinh\left(\frac{\pi \calC_3 T_3}{l}\right).\\
\end{split}.
\ee
The intrinsic-$N$ condition in  \eqref{N3eta3} can be understood as follows.
For $\frac{T_i}{l}\to 0$, one has $\text{Tr}(M_1\cdot M_2\cdot M_3)\simeq 2$.
On the other hand, by taking the limit $\frac{T_i}{l}\to \infty$,  one has
$\text{Tr}(M_1\cdot M_2\cdot M_3)\simeq \eta_3 \cdot \frac{\pi T_1}{l}
\cosh\left(\frac{\pi \calC_2 T_2}{l}\right)
\cosh\left(\frac{\pi \calC_3 T_3}{l}\right)\to -\infty$ for $\eta_3<0$,
where we have considered $\cosh\frac{\pi \calC_i T_i}{l}\simeq \sinh\frac{\pi \calC_iT_i}{l}$
and $\eta_3=2\frac{\b{\calC}_1\cdot \b{\calC}_3}{\calC_3}+
2\frac{\b{\calC}_1\cdot \b{\calC}_2}{\calC_2}+
2\frac{\b{\calC}_1*\b{\calC}_2*\b{\calC}_3}{\calC_2\,\calC_3}$.
Here one can check explicitly that $\eta_3=\text{Tr}(P_1\cdot P_2\cdot P_3)$,
with $P_i$ expressed in \eqref{PjParabolic} and \eqref{PjHyperbolic}.
Then as we tune the parameters $\frac{T_i}{l}$ continuously, there
must exist non-heating phases with $|\text{Tr}(M_1\cdot M_2\cdot M_3)|<2$.

\textit{-- 2 parabolic and 1 hyperbolic Hamiltonians}

Now let us consider the case with two parabolic and one 
hyperbolic driving Hamiltonians. 
Without loss of generality, let us choose $H_1$  and $H_2$ to be parabolic, and $H_3$
to be hyperbolic.
Based on Eqs.~\eqref{ParabolicMobius} and \eqref{HyperbolicMobius}, one has
\be
\small
\begin{split}
&\text{Tr}(M_1\cdot M_2\cdot M_3)=2
\cosh\left(\frac{\pi \calC_3 T_3}{l}\right)\\
&+2\frac{\b{\calC}_1\cdot \b{\calC}_3}{\calC_3}\cdot \frac{\pi T_1}{l}\cdot
\sinh\left(\frac{\pi \calC_3 T_3}{l}\right)\\
&+2\frac{\b{\calC}_2\cdot \b{\calC}_3}{\calC_3}\cdot \frac{\pi T_2}{l}\cdot
\sinh\left(\frac{\pi \calC_3 T_3}{l}\right)\\
&+2\,\b{\calC}_1\cdot \b{\calC}_2\cdot \frac{\pi T_1}{l}\cdot\frac{\pi T_2}{l}\cdot
\cosh\left(\frac{\pi \calC_3 T_3}{l}\right)
\\
&+2\,\frac{\b{\calC}_1*\b{\calC}_2*\b{\calC}_3}{\calC_3}\cdot \frac{\pi T_1}{l}\cdot\frac{\pi T_2}{l}\cdot
\sinh\left(\frac{\pi \calC_3 T_3}{l}\right)\\
\end{split}
\ee
As before, in the limit $\frac{T_i}{l}\to 0$, one has 
$\text{Tr}(M_1\cdot M_2\cdot M_3)\simeq 2$.
In the other limit $\frac{T_i}{l}\to \infty$,  one has
$\text{Tr}(M_1\cdot M_2\cdot M_3)\simeq \eta_3 \cdot \frac{\pi T_1}{l}
\cdot \frac{\pi T_2}{l}\cosh\left(\frac{\pi \calC_3 T_3}{l}\right)\to -\infty$ for $\eta_3<0$,
where we have considered $\cosh\frac{\pi \calC_i T_i}{l}\simeq \sinh\frac{\pi \calC_iT_i}{l}$
and $\eta_3=2 \b{\calC}_1\cdot \b{\calC}_2+
2\frac{\b{\calC}_1*\b{\calC}_2*\b{\calC}_3}{\calC_2\,\calC_3}$.
Here one can check explicitly that $\eta_3=\text{Tr}(P_1\cdot P_2\cdot P_3)$,
with $P_i$ expressed in \eqref{PjParabolic} and \eqref{PjHyperbolic}.
Apparently, as we tune the parameters $\frac{T_i}{l}$ from $0$ to $\infty$,
there exist non-heating phases with $|\text{Tr}(M_1\cdot M_2\cdot M_3)|<2$.

\newpage

\bibliography{floquet_cft}

\end{document}